\documentclass[a4paper, 11pt]{article}
\usepackage{epsfig,amsmath,amsfonts,hyperref}

\newcommand{\be}{\begin{equation}}
\newcommand{\ee}{\end{equation}}

\newcommand{\doublet}[2]{\left(\begin{array}{c}#1\\#2\end{array}\right)}
\newcommand{\triplet}[3]{\left(\begin{array}{c}#1\\#2\\#3\end{array}\right)}
\newcommand{\twobytwo}[4]{\left(\begin{array}{cc} #1&#2\\#3&#4\end{array}\right)}
\newcommand{\threebythree}[9]{\left(\begin{array}{ccc} #1&#2&#3\\#4&#5&#6\\#7&#8&#9\\\end{array}\right)}

\newcommand{\half}{\frac12}

\newcommand{\R}[4]{R^{#1\phantom{#2}#3}_{\phantom{#1}#2\phantom{#3}#4}}

\newcommand{\Rt}[4]{\widetilde{R}^{#1\phantom{#2}#3}_{\phantom{#1}#2\phantom{#3}#4}}
\newcommand{\Urm}{\mathrm{U}}
\newcommand{\su}{\mathrm{su}}

\newcommand{\SU}{\mathrm{SU}}
\newcommand{\SO}{\mathrm{SO}}

\newcommand{\Tr}{\mathrm{Tr}}

\newcommand{\Fcal}{\mathcal{F}}
\newcommand{\Jcal}{\mathcal{J}}

\newcommand{\Wcal}{\mathcal{W}}
\newcommand{\mparagraph}[1]{\paragraph{#1}\mbox{}}

\newcommand{\Zset}{\mathbb{Z}}
\newcommand{\Ncal}{\mathcal{N}}
\newcommand{\Hcal}{\mathcal{H}}
\newcommand{\Gcal}{\mathcal{G}}

\newcommand{\Srm}{\mathrm{S}}
\newcommand{\Cset}{{\,\,{{{^{_{\pmb{\mid}}}}\kern-.47em{\mathrm C}}}}}

\newcommand{\zb}{{\bar{z}}}
\newcommand{\diff}{\mathrm{d}}
\newcommand{\p}{\partial}
\newcommand{\ra}{\longrightarrow}
\newcommand{\gt}{\tilde{g}}

\newcommand{\trr}{\triangleright}
\newcommand{\id}{\mathrm{id}}

\newcommand{\comment}[1]{}

\begin{document}

\numberwithin{equation}{section}

\mbox{}

\vspace{40pt}

\begin{center}

 {\Large \bf Integrable Hopf twists, marginal deformations \\ and generalised geometry}\\
\vspace{43pt}

{\large {\mbox{{\bf Hector Dlamini$\,{}^a$} \hspace{.2cm} and \hspace{.2cm} {\bf Konstantinos Zoubos$\,{}^b$}}}}%

\vspace{.5cm}

Department of Physics, University of Pretoria\\
Private Bag X20, Hatfield 0028, South Africa

\mbox{}

and

\mbox{}

National Institute for Theoretical Physics (NITheP) \\
Gauteng, South Africa

\vspace{40pt}

{\Large \bf Abstract}

\end{center}

\vspace{.3cm}

\large

\noindent The Leigh-Strassler family of $\Ncal=1$ marginal deformations of the $\Ncal=4$ SYM theory
admits a Hopf algebra symmetry which is a quantum group deformation of the $\SU(3)$
part of the R-symmetry of the $\Ncal=4$ theory. We investigate how this quantum symmetry might be expressed on
the gravity side of the AdS/CFT correspondence. First, we discuss the twist leading to the Hopf 
algebra structure for the well-known $\beta$-deformation as well as a unitarily
equivalent theory that we call the $w$-deformation. We then show how this Hopf twist can be used to define
a star product between the three scalar superfields of these theories which encodes the deformed global
symmetry. Turning to the gravity side, we adapt this star product
to deform the pure spinors of six-dimensional flat space in its generalised geometry description. This
leads to an $\Ncal=2$ NS-NS solution of IIB supergravity. Starting from this precursor solution,
adding D3-branes and taking the near-horizon limit reproduces the dual gravitational solution to the
above theories, first derived by Lunin and Maldacena using TsT techniques. This indicates that the Hopf 
algebra symmetry can play a useful role in constructing the supergravity duals of the general Leigh-Strassler 
deformations.

\normalsize

\noindent

\vspace{3.6cm}
\noindent\rule{4.5cm}{0.4pt}

\noindent $^a$ sickmech@gmail.com 

\noindent $^b$ kzoubos@up.ac.za

\vspace{0.5cm}

\setcounter{page}{0}
\thispagestyle{empty}
\newpage

\tableofcontents

\section{Introduction}

The study of the AdS/CFT correspondence \cite{Maldacena:1997re} has provided unprecedented
insight into the behaviour of strongly-coupled quantum field theory. The additional computational power arising when
integrability \cite{Minahan:2002ve} is present has led to very precise checks of the correspondence, as well as 
a deeper understanding of how the interpolation between weak and strong coupling takes place (see \cite{Beisert:2010jr}
for an extensive review). A common theme of these studies is the existence of a large amount of symmetry in 
certain gauge theories which is not directly visible but, when identified and used to its full extent, 
can be very powerful in constraining the observables of the theory. 

 Conformal field theories play a very special role in AdS/CFT and especially in the study of integrability,
with the most fundamental example being the $\Ncal=4$ Supersymmetric Yang-Mills theory ($\Ncal=4$ SYM). An important
question is how far the power of integrability extends when considering less special theories, in particular
theories with reduced supersymmetry. In
exploring this boundary of integrability, the $\Ncal=1$ marginal deformations of $\Ncal=4$ SYM studied by Leigh
and Strassler \cite{Leigh:1995ep} play a very important role. They are superconformal theories with
superpotential
\be \label{LSW}
\Wcal_{LS}=
\kappa{\mathrm {Tr}}\left(\Phi^1[\Phi^2,\Phi^3]_{q}+ \frac h3\left((\Phi^1)^3+(\Phi^2)^3+(\Phi^3)^3\right)\right)
\ee
where $[X,Y]_q=XY-qYX$ is the $q$-deformed commutator. Beyond the special point $q=1,h=0, \kappa=g$, which
 corresponds to $\Ncal=4$ SYM, by far the best understood example is that 
of the $\beta$-deformed theory, where $q=e^{i\beta}, h=0$. This is mainly due to the fact
that the dual gravitational background was constructed by O. Lunin and J. Maldacena in 2005 \cite{Lunin:2005jy}. 
Knowledge of the dual geometry has allowed very precise checks of the integrability properties of this theory. 
If $\beta$ is taken to be real (and thus $q$ to just be a phase), integrability
on the gravity side was established through studies of semiclassical strings and the explicit 
construction of a Lax pair for the string sigma model \cite{Frolov:2005ty,Frolov:2005dj}. On the gauge theory side, 
integrability was shown at one loop in \cite{Roiban:2003dw} and 
all loops in \cite{Beisert:2005if}, where its asymptotic Bethe ansatz was also derived
from the $\Ncal=4$ SYM case by a process of twisting. On the other hand, the complex-$\beta$ deformed
theory was expected to be non-integrable since the work \cite{Berenstein:2004ys}, which showed that its
one-loop Hamiltonian does not correspond to a known integrable system. This was also explained from a
quantum group perspective in \cite{Mansson:2008xv} and demonstrated on the dual gravity side using analytic 
non-integrability techniques  applied to the Lunin-Maldacena background \cite{Giataganas:2013dha}. 
See \cite{Zoubos:2010kh} for more details and references on the integrable properties of the Leigh-Strassler 
theories. 

 In order to achieve a better understanding of the full Leigh-Strassler deformations, it is of course very 
desirable to construct the supergravity background dual to the generic $(q,h)$ deformation. A major obstacle, 
however, is that turning on the $h$ parameter generically leaves only one $\Urm(1)$ unbroken out of the $\SU(4)$ 
R-symmetry of the $\Ncal=4$ theory. The corresponding dual background will then also only have
one $\Urm(1)$ isometry, the dual to the unbroken R-symmetry. This is in contrast to the $\beta$-deformation,
which breaks the $\SU(4)$ to its (Cartan)  $\Urm(1)^3$ subgroup. The presence of the additional $\Urm(1)\times\Urm(1)$
is what permits application of the TsT techniques of \cite{Lunin:2005jy}, since they depend on T-dualising along 
isometry directions. The TsT method is thus not available in the general case. One way to proceed is by 
attempting to solve the supergravity 
equations order-by-order in the deformation. This was done in \cite{Aharony:2002hx}, guided by the 
group-theory structure to turn on the additional fields required for the deformation, and imposing the 
supersymmetry variation equations to maintain the required amount of supercharges. This led to the solution to second 
order in the generic deformation. Although this approach demonstrated the existence of these deformations,
as required by AdS/CFT, it could not by its perturbative nature lead to the \emph{exact} solution. 

 A different approach to constructing the dual to the pure $h$-deformation was followed in \cite{Kulaxizi:2006zc},
which used noncommutative geometry techniques within the framework of Seiberg and Witten \cite{Seiberg:1999vs}. 
This led to a solution of the IIB equations up to third order in the deformation parameter. However, continuation 
to higher orders was  obstructed  by ambiguities related to non-associativity of the star product 
introduced in \cite{Kulaxizi:2006zc}.

On the gauge theory side, a better understanding of the symmetries of the Leigh-Strassler theories
 was achieved in \cite{Mansson:2008xv}. There it was shown that turning
on the $q$ and $h$ parameters in (\ref{LSW}) does not actually break the  
$\SU(3)\times\Urm(1)$ R-symmetry of $\Ncal=4$ SYM (the subgroup of $\SU(4)$ which is manifest in $\Ncal=1$
superspace notation). Rather, the $\SU(3)$ part is \emph{deformed} to a Hopf algebra, which we will
call $\widetilde{\SU(3)}_{q,h}$ in this work.
Using the RTT relations of Faddeev-Reshetikhin-Takhtajan \cite{FRT90}, the work \cite{Mansson:2008xv}
derived the commutation relations of this 
Hopf algebra, which was shown to be a global symmetry of the Leigh-Strassler-deformed superpotential (\ref{LSW}).

 Clearly, the fact that the $\SU(3)$ is deformed, not broken, means that the $(q,h)$-deformations have
more symmetry than is naively visible, as long as one is willing to enlarge one's toolkit beyond Lie
algebras to include Hopf-algebraic symmetries. If present also on the gravity side, this additional symmetry
can potentially be used to facilitate the construction of the dual background. However, there were two 
immediate obstacles to implementing this programme. Firstly, the Hopf algebra of \cite{Mansson:2008xv}
is \emph{non-quasitriangular} for generic $(q,h)$\footnote{See appendix \ref{Hopfappendix} for 
the definition of quasitriangular Hopf algebras.}, which can be expected to lead to problems with associativity
similar to those encountered in \cite{Kulaxizi:2006zc}. Secondly, since we are dealing with a quantum symmetry, 
it was not immediately clear how to make it manifest on the gravity side, where one of course expects to
find a smooth, commutative geometry. 

 In this work we will take some first steps towards making the Hopf-algebraic symmetry visible on the 
gravity side. To avoid issues with associativity, we will focus on two special cases admitting a 
quasitriangular structure. The first is the well-known $\beta$-deformation, which we will revisit from
the viewpoint of the Hopf-algebraic symmetry above. The second case we will consider is derived
 by taking $(q,h)=(1\!+\!w,w)$ in the superpotential above. We will call this theory the $w$-deformation 
of $\Ncal=4$ SYM. This theory is known to be unitarily equivalent to the $\beta$ deformation \cite{Bundzik:2005zg,
Bork:2007bj,Mansson:2008xv} and is thus not a new
case. However, we will argue that considering it on its own merits provides additional insight into 
the construction.  As for the second problem, we propose that a suitable framework within which to appreciate 
the consequences of the Hopf algebra symmetry on the gravity side is that of generalised geometry.  

More specifically, after a study of the (all-orders in the deformation parameter) Hopf structure of the 
$\beta$ and $w$-deformed theories, we will focus on the first order structure, which will define a 
(parameter-dependent) non-commutativity 
matrix. This will be used to define a star product between the superfields which encodes the deformation. 
 We will then propose that
the same matrix can be used on the gravity side to define a \emph{non-anticommutative} star product
on the space of generalised forms. Deforming the pure spinors of $\Cset^3$ with this product
will result in a deformed NS-NS geometry. Considering the near-horizon limit of D3-branes on this 
geometry will lead us to the full dual geometry, which is just the LM background \cite{Lunin:2005jy} 
in the coordinate system most adapted to each of the two theories. 

Apart from \cite{Mansson:2008xv}, this work draws inspiration mainly from \cite{Kulaxizi:2006zc}, 
which (as mentioned above) used non-commutativity techniques
to construct the dual to the pure $h$-deformation up to third order in the deformation parameter, and
\cite{Halmagyi:2007ft}, which discussed the generalised geometry description of the NS-NS precursor
of the real-$\beta$ background. 

This article is organised into two main parts, which can be read independently. Sections 
\ref{Marginal}-\ref{Noncommutativity} are concerned with the gauge theory side of the $\beta$ and 
$w$-deformed theories,
focusing on its Hopf algebra structure and rewriting it in terms of a star product. Readers who
are mainly interested in  geometric aspects can skip forward to sections
\ref{GeneralisedGeometry}-\ref{Metric}, which are concerned with the generalised geometry structure of
the theory and how it can be employed to construct the dual background.

\mbox{}

\mparagraph{Methodology}

All symbolic computations in this work have been performed with GPL Maxima \cite{maxima}. The relevant worksheet 
has been attached to the arXiv submission as ancillary material. 

\mbox{}

\mparagraph{Note Added} \mbox{}

In the first version of this article, it was incorrectly claimed that the $w$-deformed
IIB solution was truly different to the $\beta$-deformed one, rather than just written in a coordinate
system more adapted to the $w$-deformation. We thank our colleagues, especially M. Kulaxizi and S. van Tongeren,
for comments that clarified the situation.

\section{Marginal deformations of $\Ncal=4$ SYM} \label{Marginal}

A major milestone in the study of superconformal 4-dimensional gauge theory was the work of 
Leigh and Strassler \cite{Leigh:1995ep}. They established that, beyond the $\Ncal=4$ SYM
fixed line parametrised by the complexified gauge coupling, the $\Ncal=4$ SYM superpotential can
also be marginally deformed along two directions, corresponding to the complex parameters $q$ and $h$ in (\ref{LSW}),
which preserve $\Ncal=1$ supersymmetry. The manifold of $\Ncal=1$ finite Leigh-Strassler theories can 
thus be parametrised by a function $f(g,\kappa,q,h)=0$. This function is not known in general, but perturbative
expressions are known (see e.g. \cite{Aharony:2002tp,Freedman:2005cg,Penati:2005hp,Rossi:2005mr}). There has been considerable recent progress in 
characterising the conformal manifold of $\Ncal=1$ SYM theories \cite{Kol:2010ub,Asnin:2009xx,Green:2010da},
of which the marginal deformations of $\Ncal=4$ SYM are a special case. 

The link between the Leigh-Strassler deformations and non-commutativity was perceived early on, starting
with \cite{Berenstein:2000ux}, which studied the non-commutative geometry of their moduli spaces of 
vacua. In the context of the twistor string \cite{Witten:2003nn}, it was shown in \cite{Kulaxizi:2004pa}
that the amplitudes in these theories can be computed, to first order in the deformation parameter, via 
a simple \emph{non-anticommutative} star product between the fermionic coordinates of supertwistor space. 
Non-associativity hindered the extension of this construction to higher orders, which was however achieved in
\cite{Gao:2006mw} for the special case of the real-$\beta$ deformation where associativity is present. 

 In the special case of the real-$\beta$ deformations, a (non-commutative) Moyal star product taking the
$\Ncal=4$ SYM theory to the deformed theory was
introduced by Lunin and Maldacena \cite{Lunin:2005jy}, inspiring their construction of the dual geometry 
using a combination of T-duality, shifting along a $\Urm(1)$ direction and another T-duality. The simple
relation between the $\Ncal=4$  SYM and $\beta$-deformed theory provided by this star product facilitated the
construction of its Bethe ansatz \cite{Beisert:2005if} and paved the way for many detailed studies and
checks of AdS/CFT in this less supersymmetric setting. Reviews of these
developments from an integrability perspective can be found in \cite{Zoubos:2010kh} and \cite{vanTongeren:2013gva}.

 There have been comparatively few studies of the Leigh-Strassler theories beyond the real-$\beta$ case.
In \cite{Mansson:2007sh} it was shown that the non-integrable complex-$\beta$ theory has a one-loop
integrable $\SU(3)$ subsector, which was further studied in \cite{Puletti:2011hx} in the context of string
motion on the dual geometry. Going beyond the $\beta$-deformation, the work \cite{Minahan:2011dd} studied 
higher-loop anomalous dimensions in a very interesting corner of the Leigh-Strassler conformal manifold, where 
$\kappa$ and $h$ in (\ref{LSW}) are scaled such that only the cubic second term remains. Since this theory
is in a sense maximally distant from the undeformed case, it is expected that it will not have a good supergravity
dual. Rather, stringy corrections are likely to play an important role. The higher-loop behaviour of this theory
as well as the complex-$\beta$ case was further studied in \cite{Minahan:2011bi}. 

The one-loop integrability of the general Leigh-Strassler theories
was studied in \cite{Bundzik:2005zg,Mansson:2007sh}, leading up to the work \cite{Mansson:2008xv} 
which uncovered the Hopf algebra symmetry underlying the superpotential of the general deformations.
The main goal of the latter work was to understand the interplay between finiteness and integrability
of the Leigh-Strassler theories. It was found that although the $\widetilde{\SU(3)}_{q,h}$ Hopf algebra 
structure is present for the general, non-integrable, $(q,h)$-deformed
theory, integrability is only expected to arise when the Hopf algebra is \emph{quasitriangular}.
As we will discuss in the next section, this is the case when the $R$-matrix defining the Hopf algebra
can be derived from that of the undeformed $\Ncal=4$ theory by a transformation called a Hopf (or Drinfeld) twist. 

 Generically, however, the Hopf algebra was found to be non-quasitriangular. As discussed in \cite{Mansson:2008xv},
in that case the requirement of associativity (which is a defining property of Hopf algebras) necessarily
imposes conditions on the Hopf algebra generators at cubic and higher level which do not follow from the (quadratic)
commutation relations. A potential consequence of this is the trivialisation of the higher-order 
algebra (the additional relations appearing at each higher level eventually setting all products of generators
above some level to zero). However, in \cite{Mansson:2008xv} it was shown that the cubic quantum determinant 
is non-trivial 
and central. Furthermore, in \cite{ManssonPBW} a Poincar\'e-Birkhoff-Witt basis for this algebra was
constructed for the special case where $q=\bar{q}$, $h=\bar{h}=0$. This all points towards the existence 
of a consistent and
non-trivial higher-order structure for the $\widetilde{\SU(3)}_{q,h}$ algebra for general $q$ and $h$,
which however has not yet been uncovered. 

 As already mentioned, in this work we will sidestep the issue of non-associativity by focusing on two 
special cases, the $\beta$ and $w$-deformations where the Hopf algebra is quasi-triangular. 
The theories are equivalent, but we find it useful to study the $w$-deformation independently both as a 
consistency check and because the $w$-deformed frame might provide a better starting point for generalisations to
generic deformations. In particular,  the $\Urm(1)\times \Urm(1)$ symmetry of the $\beta$-deformation is obscured
in the $w$-deformation, which means that to some extent we will be working as if that symmetry is not there.
We start by reviewing the Hopf structure of the $\beta$-deformed theory in the
next section before moving on to the $w$ deformed theory in the section after that. 

 In the above we reviewed how  both non-commutative and non-anticommutative star products have been proposed
in order to deform the $\Ncal=4$ SYM theory to the Leigh-Strassler theories in different contexts. In this work, 
we  will first use the Hopf structure to define a \emph{non-commutative} star product on the gauge theory side of the 
AdS/CFT correspondence, however will later use the same non-commutativity matrix  to construct a 
\emph{non-anticommutative} star product on the gravity side, which will act on the space of polyforms in the
context of generalised geometry.

\subsection{Hopf structure of the Leigh-Strassler theories}

In \cite{Mansson:2008xv} it was shown that the superpotential (\ref{LSW}) is invariant under
a Hopf algebra which can be thought of as a deformation of $\SU(3)$. We will call this Hopf
algebra $\widetilde{\SU(3)}_{q,h}$ or sometimes just $\Hcal$. The algebra is defined through the RTT relations \cite{FRT90}
using the following $R$-matrix, related in a simple way to the one-loop Hamiltonian of the theory in the $\SU(3)$ sector,
 acting on $\Hcal\otimes \Hcal$:\footnote{We order the basis on $\Hcal\otimes \Hcal$ as $\{11,12,13,21,22,23,31,32,33\}$, so we read off  $R^{11}_{\;11}=(1+q\bar{q}-h\bar{h})/(2d^2),R^{11}_{\;23}=-2\bar{h}/(2d^2)$ etc.} 
\small  
\be \label{qhRmatrix}
R=\frac{1}{2d^2}
\begin{pmatrix}
1\!+\!q\bar{q}\!-\!h\bar{h}&\hspace{-.3cm}0&\hspace{-.3cm}0&\hspace{-.3cm}0&\hspace{-.3cm}0&\hspace{-.3cm}\!-\!2\bar{h}&\hspace{-.3cm}0&\hspace{-.3cm}2\bar{h}q&\hspace{-.3cm}0\cr 
0&\hspace{-.3cm}2\bar{q}&\hspace{-.3cm}0&\hspace{-.3cm}1\!-\!q\bar{q}\!+\!h\bar{h}&\hspace{-.3cm}0&\hspace{-.3cm}0&\hspace{-.3cm}0&\hspace{-.3cm}0&\hspace{-.3cm}2 h\bar{q}\cr 
0&\hspace{-.3cm}0&\hspace{-.3cm}2q&\hspace{-.3cm}0&\hspace{-.3cm}\!-\!2h&\hspace{-.3cm}0&\hspace{-.3cm}q\bar{q}\!+\!h\bar{h}\!-\!1&\hspace{-.3cm}0&\hspace{-.3cm}0\cr 
0&\hspace{-.3cm} q\bar{q}\!+\!h\bar{h}\!-\!1&\hspace{-.3cm}0&\hspace{-.3cm}2q&\hspace{-.3cm}0&\hspace{-.3cm}0&\hspace{-.3cm}0&\hspace{-.3cm}0&\hspace{-.3cm}\!-\!2h\cr 
0&\hspace{-.3cm}0&\hspace{-.3cm}2\bar{h}q&\hspace{-.3cm}0&\hspace{-.3cm}1\!+\! q\bar{q}\!-\!h\bar{h}&\hspace{-.3cm}0&\hspace{-.3cm}\!-\!2\bar{h}&\hspace{-.3cm}0&\hspace{-.3cm}0\cr 
2h\bar{q}&\hspace{-.3cm}0&\hspace{-.3cm}0&\hspace{-.3cm}0&\hspace{-.3cm}0&\hspace{-.3cm}2 \bar{q}&\hspace{-.3cm}0&\hspace{-.3cm}1\!-\!q\bar{q}\!+\!h\bar{h}&\hspace{-.3cm}0\cr 
0&\hspace{-.3cm}0&\hspace{-.3cm}1\!-\!q \bar{q}\!+\!h\bar{h}&\hspace{-.3cm}0&\hspace{-.3cm}2h\bar{q}&\hspace{-.3cm}0&\hspace{-.3cm}2\bar{q}&\hspace{-.3cm}0&\hspace{-.3cm}0\cr 
\!-\!2 h&\hspace{-.3cm}0&\hspace{-.3cm}0&\hspace{-.3cm}0&\hspace{-.3cm}0&\hspace{-.3cm}q\bar{q}\!+\!h\bar{h}\!-\!1&\hspace{-.3cm}0&\hspace{-.3cm}2q&\hspace{-.3cm}0\cr
 0&\hspace{-.3cm}\!-\!2\bar{h}&\hspace{-.3cm}0&\hspace{-.3cm}2\bar{h}q&\hspace{-.3cm}0&\hspace{-.3cm}0&\hspace{-.3cm}0&\hspace{-.3cm}0&\hspace{-.3cm}1\!+\!q\bar{q}\!-\!h\bar{h}\cr 
\end{pmatrix}
\ee
\normalsize
 Here $2d^2=1+\bar{q}q+\bar{h}h$. We
note that this $R$-matrix is cyclic, $R^{ij}_{kl}=R^{(i+1)(j+1)}_{(k+1)(l+1)}\!=\!R^{(i\!-\!1)(j\!-\!1)}_{(k\!-\!1)(l\!-\!1)}$,
which guarantees compatibility of the Hopf algebra structure with the gauge theory trace.

Another important property of $R$ is that it satisfies a triangular-type condition:
\be \label{triangular}
R_{21}=R^{-1}_{12} \qquad \left(\text{in index notation:}\quad R^{nm}_{ \;\;ji} R^{ij}_{kl}=\delta^m_k\delta^n_l\right)\;.
\ee
The significance of this condition will be seen momentarily. Despite the property (\ref{triangular}),
the $R$-matrix (\ref{qhRmatrix}) does \emph{not} in general satisfy the Yang-Baxter equation. This
happens only for certain special cases of the parameters $q$ and $h$, which will be our focus in this
work.\footnote{The triangularity condition is usually thought of as a special case of the quasitriangularity
condition, so the fact that it is satisfied by the non-quasitriangular $R_{q,h}$ might appear odd. Note, however,
that quasitriangularity requires a Hopf twist (satisfying the cocycle condition), while the (still-to-be-constructed)
twist leading to $R_{q,h}$ is not expected to be a Hopf twist. It is likely that the appropriate context to understand
$R_{q,h}$ is in terms of a triangular quasi-Hopf twist. A similar situation has been described in e.g. \cite{Young:2008zm}.}

For our present purposes we are not interested in the Hopf algebra $\Hcal$ itself (for which we refer
to \cite{Mansson:2008xv}), but rather in the structures that it induces on the \emph{Hopf algebra module},
that is the representation space on which it acts. The structure is that of a \emph{quantum plane}, defined
as 
\be \label{qplane}
z^j z^i=R^{ij}_{\;kl} z^k z^l\;.
\ee
Here the indices $i,j,\ldots$ range from $1$ to $3$.
Clearly the coordinates $z^i$ are non-commuting when $R$ is non-trivial. 
They reduce to the commuting coordinates of $\Cset_3$ when $(q,h)=(1,0)$
and thus $R=I\otimes I$. In \cite{Mansson:2008xv} it was shown that the Hopf algebra $\Hcal$ acts as a symmetry
algebra on this quantum plane, i.e. the action of $\Hcal$ rotates the $z^i$ around while preserving the condition
(\ref{qplane}). 

The physical relevance of this 
quantum plane structure arises by identifying the deformed $\Cset_3$ of the quantum plane with the open-string
geometry of the six-dimensional
space transverse to the stack of D3-branes defining the theory. Since the three scalar fields $\Phi^i$ probe
the geometry of this space, the fact that the transverse space is a non-commuting quantum plane leads to the 
non-commutative moduli space of vacua explored in \cite{Berenstein:2000ux}, which takes precisely the form (\ref{qplane}). However, the study of 
\cite{Mansson:2008xv} showed that the $\widetilde{\SU(3)}_{q,h}$ quantum symmetry acting on this 
quantum plane does not only appear at 
the level of the moduli space, but is an actual global symmetry of the superpotential (\ref{LSW}). 

 An important comment about the quantum plane relations (\ref{qplane}) with the $R$ given in (\ref{qhRmatrix}) 
is that they are \emph{cyclic}, i.e.
invariant under cyclic permutations of the indices. This is unlike the more well-known quantum planes
related to the usual $\SU(3)_q$ deformations of $\SU(3)$, which are \emph{ordered} as $i<j$. The cyclicity
of the quantum plane relations is crucial in ensuring compatibility of the noncommutative structure with
the non-abelian structure of the scalar fields, which are of course also $N\times N$ matrices entering 
in the gauge theory trace \cite{Mansson:2008xv}.

 Before proceeding to the special cases which will occupy us in this work, let us see why the condition
(\ref{triangular}) above is necessary for the consistency of the quantum plane, by using the definition
(\ref{qplane}) twice:
\be
z^j z^i=R^{ij}_{kl}z^k z^l=R^{ij}_{kl} R^{lk}_{mn} z^m z^n= \delta^j_m \delta^i_n z^m z^n=z^j z^i\;.
\ee

\section{The $\beta$-deformed theory and its Hopf structure} \label{Hopfbeta}

As a warm-up, and to set the stage, in this section we will review the Hopf structure of 
the real $\beta$-deformed theory. 
This is a well-known integrable deformation of $\Ncal=4$ SYM, with superpotential
\be \label{qdeformedW}
\Wcal=\kappa \Tr[\Phi^1\Phi^2\Phi^3-q \Phi^1\Phi^3\Phi^2]
\ee
where $q=e^{i\beta}$, with $\beta$ a real parameter. The corresponding $R$-matrix 
\cite{Beisert:2005if, Mansson:2008xv} is diagonal and obtained by setting $(q,h)=(q,0)$ in (\ref{qhRmatrix}),
with $\bar{q}=1/q$:\footnote{The full, spectral parameter-dependent 
$R$-matrix can be found in \cite{Beisert:2005if}. In (\ref{qRmatrix}) we have
already taken the limit of infinite spectral parameter  
where the link to the underlying quantum symmetry becomes apparent.}
\be \label{qRmatrix}
R_q=
\begin{pmatrix}
1&0&0&0&0&0&0&0&0\cr 
0&q^{-1}&0&0&0&0&0&0&0\cr 
0&0&q&0&0&0&0&0&0 \cr 
0&0&0&q&0&0&0&0&0\cr 
0&0&0&0&1&0&0&0&0\cr
0&0&0&0&0&q^{-1}&0&0&0\cr 
0&0&0&0&0&0&q^{-1}&0&0\cr 
0&0&0&0&0&0&0&q&0 \cr 
0&0&0&0&0&0&0&0&1\cr
\end{pmatrix} \;.
\ee
This $R$-matrix defines a quantum-plane structure on a deformed $\Cset_3$ through (\ref{qplane}).
 The $\Ncal=4$ theory corresponds to $q=1$ in (\ref{qplane}), with the $R$-matrix being simply $R_1=I\otimes I$ and 
commutativity restored. In that case the symmetry leaving the plane invariant is just the usual Lie-algebraic
$\SU(3)$. 

A crucial feature of  $R_q$ is that it can be related to the undeformed 
$R$-matrix $R_1=I\otimes I$ through a twisting relation  
\be \label{twisting}
R_q=F_{q,21}\cdot (I\otimes I) \cdot F_q^{-1}
\ee
with the matrix $F_q$ being simply
\be \label{qTwist}
F_q=
\begin{pmatrix}
1&0&0&0&0&0&0&0&0\cr 
0&q^\half&0&0&0&0&0&0&0\cr 
0&0&q^{-\half}&0&0&0&0&0&0 \cr 
0&0&0&q^{-\half}&0&0&0&0&0\cr 
0&0&0&0&1&0&0&0&0\cr
0&0&0&0&0&q^\half&0&0&0\cr 
0&0&0&0&0&0&q^\half&0&0\cr 
0&0&0&0&0&0&0&q^{-\half}&0 \cr 
0&0&0&0&0&0&0&0&1\cr
\end{pmatrix} \;.
\ee
In general, a matrix $F$ relating the undeformed to the deformed $R$-matrix via the generalised 
similarity transformation (\ref{twisting}), is known as a \emph{twist}. 
In order for it to be a \emph{Hopf twist} (also known as a Drinfeld twist), it needs to satisfy the cocycle identity
\be \label{cocycle}
(F\otimes 1)(\Delta \otimes \id) F=(1 \otimes F)(\id\otimes \Delta) F
\ee
with $\Delta$ being the coproduct of the undeformed Hopf algebra \cite{Drinfeld90}. This identity holds for 
the twist $F_q$ above.  
To show this, we can use the fact that the twist $F_q$ (being abelian) can be written 
in terms of the Cartan generators of $\SU(3)$  \cite{Reshetikhin90,Matsumoto:2014nra,vanTongeren:2015soa,vanTongeren:2015uha}. One way to express it is
\be
F_q=e^{i\frac{\beta}2 H_1\wedge H_2}\;,\quad \quad \text{where}\quad H_1=\threebythree{1}{0}{0}{0}{-1}{0}{0}{0}{0}\;,
\;\;H_2=\threebythree{0}{0}{0}{0}{1}{0}{0}{0}{-1}\;.
\ee
Since the usual Lie algebraic coproduct
\be
\Delta(H_i)=H_i\otimes I + I \otimes H_i
\ee
simply exponentiates (again because the deformation is abelian), we find
\be
\begin{split}
(\Delta \otimes \id) F_q&=(\Delta \otimes \id) e^{i\frac\beta 2 (H_1\otimes H_2-H_2\otimes H_1)}=
e^{i\frac\beta 2 (\Delta(H_1)\otimes H_2-\Delta(H_2)\otimes H_1)}\\
&=e^{i\frac\beta 2( H_1\otimes 1\otimes H_2 + 1\otimes H_1 \otimes H_2-H_2\otimes 1 \otimes H_1 - 1\otimes H_2 \otimes H_1)}\\
&=e^{i\frac\beta 2( H_1\otimes 1\otimes H_2  -H_2\otimes 1 \otimes H_1)} e^{i\frac\beta 2(1\otimes H_1 \otimes H_2- 1\otimes H_2 \otimes H_1)}
=F_{q,13}F_{q,23}\;.
\end{split}
\ee
We can similarly show that $(\id\otimes \Delta)F_q=F_{q,13}F_{q,12}$. Thus (\ref{cocycle}) becomes
\be
F_{q,12}F_{q,13}F_{q,23}=F_{q,23}F_{q,13}F_{q,12}
\ee
which can be straightforwardly checked.\footnote{Writing this expression as an YBE is for illustration only, since
of course $F_q$ here is abelian so any ordering of terms would produce the same result.} The counital 
condition $(\epsilon\otimes \text{id})F=1$ is automatic, since the counit satisfies $\epsilon(1)=1$ and
$\epsilon(X)=0$ for any other algebra element. 
 It is also easy to show that $R_q=F_{21}F^{-1}_{12}$ satisfies the quasitriangular identities
\be
(\Delta_F\otimes\mathrm{id}) R=R_{13}R_{23}\;,\quad (\mathrm{id}\otimes \Delta_F)R=R_{13}R_{12}\;,
\ee
as well as
\be
\tau\circ \Delta_F=R \Delta_F R^{-1}\;.
\ee
Here $\Delta_F$ denotes the twisted coproduct
\be \label{DeltaF}
\Delta_F(X)=F\Delta(X) F^{-1}\;,
\ee
 although in practice it
turns out to be equal to the undeformed coproduct when acting on $H_{1,2}$. As discussed in 
appendix \ref{Hopfappendix}, the quasitriangular identities imply the Yang-Baxter equation for $R$. 

For later use, let us write down the twisting relation (\ref{twisting}) explicitly, along with 
a similar (inverse) relation satisfied by $R_q$ which can straightforwardly be checked:
\be \label{RFrels}
R^{ij}_{rs}=(R^{-1})^{ji}_{sr}=F^{ji}_{lk} (F^{-1})^{kl}_{rs} \quad \text{and}\quad 
R^{ji}_{sr}=(R^{-1})^{ij}_{rs}=(F^{-1})^{ji}_{lk} F^{kl}_{rs}
\ee
Finally, let us note that (\ref{twisting}) trivially implies the triangularity condition (\ref{triangular}):
\be \label{triangularbeta}
(R_{21}R)^{ij}_{\;kl}=R^{ji}_{\;nm} R^{mn}_{\;kl}=F^{ij}_{\;pq} (F^{-1})^{qp}_{\;nm} F^{nm}_{\;sr} (F^{-1})^{rs}_{\;kl}=\delta^i_k \delta^j_l\;.
\ee
Unlike the general case of the $(q,h)$ $R$-matrix, here the triangularity condition follows from 
a twist satisfying the cocycle condition. Since the starting point $I\otimes I$ is a 
quasitriangular Hopf algebra, so is the deformed algebra $\widetilde{SU}_q$. The condition (\ref{triangularbeta})
tells us that it is actually a triangular Hopf algebra. (This is not the case for the general $(q,h)$ case
since the twist from $I\otimes I$ is not expected to be a Hopf twist). 

In conclusion, from our perspective the (planar) $\beta$-deformed theory is nothing but a Hopf-twisted 
version of the $\Ncal=4$ SYM theory. One thus expects its properties to follow simply from those of the 
undeformed theory, which has been shown to be the case, at least as far as the asymptotic Bethe ansatz
is concerned. In particular, it is integrable, with
its Bethe ansatz written down in \cite{Beisert:2005if} by twisting 
(adding appropriate phases to) the $\Ncal=4$ SYM Bethe ansatz.\footnote{There are also significant
differences between $\Ncal=4$ SYM and the $\beta$-deformed theory. For instance, wrapping corrections 
arise at lower loop orders in the latter theory compared to the undeformed one \cite{Fiamberti:2008sm,Fokken:2013mza}.
Certainly the non-planar behaviour of the two theories is not expected to be related in such a simple way.} 

 From a Hopf-algebra perspective, the above construction is slightly trivial since the twisting matrix is
diagonal, which also leads to a diagonal deformed $R$-matrix. To better appreciate the formalism, it is
important to apply it to a non-diagonal case, which is what we will do in the next section.

\section{The $w$-deformed theory and its Hopf structure} \label{wDeformation}

Having reviewed the Hopf structure of the $\beta$-deformation, we will now focus on another one-loop
integrable Leigh-Strassler 
deformation, first identified in \cite{Bundzik:2005zg} and further studied in \cite{Bork:2007bj,Mansson:2008xv}. 
It arises from the choice
\be \label{qhrcase}
q=(1+w) \quad \mbox{and}\quad h=w  
\ee
in the superpotential (\ref{LSW}), with $w$ a real parameter\footnote{The parameter $w$ was denoted $\rho$ in
\cite{Bundzik:2005zg} and $r$ in \cite{Mansson:2008xv}. We relabelled it here to avoid potential confusion 
with the $\rho$-deformation (an alternative notation for the pure-$h$ deformation), as well as the
classical $r$-matrix and the AdS radial parameter, both of which make an appearance later on. We also 
note that this deformation belongs to a more general class of integrable theories 
parametrised by phases \cite{Bundzik:2005zg,Mansson:2008xv}, which we set to zero in this work for simplicity.}. Explicitly,
the superpotential of the $w$-deformed theory is
\be \label{wdefW}
\Wcal=\kappa \Tr\left[\Phi^1\Phi^2\Phi^3-(1+w) \Phi^1\Phi^3\Phi^2 + \frac{w}{3}\left((\Phi^1)^3+(\Phi^2)^3+(\Phi^3)^3\right)\right]
\ee
We see that the $w$-deformation is a very special combination of a $q$ and $h$ deformation.
 
 As discussed in \cite{Bork:2007bj}, the $w$-deformation is actually unitarily equivalent
to the $\beta$-deformation. More precisely, the above superpotential can be derived from the $\beta$-deformed
one (\ref{qdeformedW}) through a field redefinition $\Phi^i\ra (T^\dag)^i_{\;j} \Phi^j$, with $T$ the $\SU(3)$ matrix\footnote{To 
obtain precisely (\ref{wdefW}), the parameter $q$ needs to be rescaled as in (\ref{qrescaling}) and the overall factor 
$\kappa$ as $\kappa=(1+w e^{\frac{\pi i}3})$, which gives $|\kappa|^2=1+w+w^2$. This results in the conformality condition 
$|\kappa|^2=g^2$ holding for a different value of $g$.}
\be \label{Tmatrix}
T=-\frac i{\sqrt{3}}\begin{pmatrix} 1&1&1
\cr 1 & e^{i\frac{2\pi}{3}} & e^{-i\frac{2\pi}{3}}
\cr  1 & e^{-i\frac{2\pi}{3}} & e^{i\frac{2\pi}{3}}\cr
\end{pmatrix} \;.
\ee
This of course implies that all on-shell quantities will be equal between the two theories, while off-shell 
quantities like correlation functions can be related by applying the field redefinition. As
we will see, this difference will manifest itself in a different coordinate system being more natural to 
describe the dual geometry of the $w$-deformation. For our purposes, the importance of the field redefinition
is that it obscures the $\Urm(1)\times \Urm(1)$ symmetry present in the $\beta$-deformed superpotential and
thus allows us to test our methods in a setup that is closer to the generic case.

Like all of the $(q,h)$ deformations, the superpotential (\ref{wdefW}) admits two very important $\Zset_3$ symmetries, 
namely $\Phi^k\ra \Phi^{k+1}$ and $\Phi^k\ra e^{2\pi i k/3}\Phi^k$. The first of these $\Zset_3$'s will be
used extensively in the following to reduce the number of independent expressions under consideration. 

The $w$-deformed $R$-matrix can be found by making the above substitutions in the full $(q,h)$-deformed
$R$-matrix studied in \cite{Bundzik:2005zg,Mansson:2008xv}. It is: 
\be \label{wRmatrix}
R_w=\frac{1+w}{1+w+w^2}
\begin{pmatrix}1&0&0&0&0&-{{w}\over{1+w}}&0&w&0\cr 0&1&0&-{{w}\over{1+w}}
 &0&0&0&0&w\cr 0&0&1&0&-{{w}\over{1+w}}&0&w&0&0\cr 0&w&0&1&0&0&0&0&-
 {{w}\over{1+w}}\cr 0&0&w&0&1&0&-{{w}\over{1+w}}&0&0\cr w&0&0&0&0&1&0
 &-{{w}\over{1+w}}&0\cr 0&0&-{{w}\over{1+w}}&0&w&0&1&0&0\cr -{{w
 }\over{1+w}}&0&0&0&0&w&0&1&0\cr 0&-{{w}\over{1+w}}&0&w&0&0&0&0&1\cr 
\end{pmatrix}.
\ee
In \cite{Mansson:2008xv} it was shown that this $R$-matrix can be obtained from the $\beta$-deformed
$R$-matrix (\ref{qRmatrix}) by a twist taking the form $\Fcal =T  \otimes T$, which is nothing but a change
of basis in the algebra. The relation between the couplings is 
\be \label{qrescaling}
q=\frac{1+2w e^{-\frac{\pi i}{3}}+w^2e^{-\frac{2\pi i}{3}}}{1+w+w^2}.
\ee
So, as already discussed, we expect the algebraic structure of the two theories
to be equivalent, and in particular, unlike the full $(q,h)$-deformed $R$-matrix, 
$R_w$ can be seen to satisfy the Yang-Baxter Equation, which implies that the corresponding one-loop planar 
Hamiltonian is integrable. Although the superpotential (\ref{wdefW}) might appear to produce different types
of interactions compared to the $\beta$-deformed case, in  \cite{Bundzik:2005zg} it was shown that appropriate
redefinitions bring the $w$-deformed spin-chain Hamiltonian to precisely the form of the $\beta$-deformation. So
we certainly do not expect any difference between the theories as far as integrability is concerned. 

 The main motivation to study the $w$-deformation independently of the $\beta$-deformation is that, being
superficially more complicated, it can potentially provide insights into the generic algebraic structure 
of the Leigh-Strassler deformation  and the way this structure is expressed on the gravity side that are harder to see
directly in the $\beta$-deformation. 

The matrix $R_w$ can be conveniently written in terms of the shift matrix $U$ and its square $V$:
\be \label{shift}
U=\threebythree{0}{0}{1}{1}{0}{0}{0}{1}{0}\,, \quad V:=U^2=\threebythree{0}{1}{0}{0}{0}{1}{1}{0}{0}\,,\quad U^3=I\;.
\ee
Using these matrices, we can express $R_w$ as:
\be \label{Rw}
R_w=\frac{1+w}{1+w+w^2}\left[I\otimes I + w\, U\otimes V -\frac{w}{1+w} V\otimes U\right]\;.
\ee
Note that $[U,V]=0$, so these matrices define an abelian subsector of $\SU(3)$. This is as expected,
since the $w$-deformed $R$-matrix is equivalent to the $\beta$-deformed $R$-matrix.

Let us now recover the $R$-matrix in (\ref{wRmatrix}) via a twist from the undeformed $R$-matrix of $\Ncal=4$ SYM.
A suitable such twist is:\footnote{This turns out to be a more apt choice than the one presented in \cite{Mansson:2008xv}, which did not smoothly reduce to the trivial twist as $w=0$.}
\be \label{wtwistmatrix}
F_w=\tilde{C}\begin{pmatrix}
1+w&0&0&0&0&w&0&0&0\cr 0&1+w&0&w&0&0&0&0&0\cr 0&0&1+w&0&w&
 0&0&0&0\cr 0&0&0&1+w&0&0&0&0&w\cr 0&0&0&0&1+w&0&w&0&0\cr 0&0&0&0&0&1+w
 &0&w&0\cr 0&0&w&0&0&0&1+w&0&0\cr w&0&0&0&0&0&0&1+w&0\cr 0&w&0&0&0&
 0&0&0&1+w\cr
 \end{pmatrix}
\ee
which can be written in a more compact form as
\be \label{wtwist}
F_w=\tilde{C}\big[(1+w) I\otimes I +w V\otimes U\;\big].
\ee
Here $\tilde{C}$ is a normalisation constant. It is in principle an arbitrary function of $w$ (chosen to reduce
 to $1$ in the  $w\ra 0$ limit) as it cancels out in the
twisted $R$-matrix (\ref{twisting}). However, below we will specify a choice that is particularly natural and
will also make $F_w$ into a counital twist.
 
We will also be interested in the inverse of $F_w$, which is
\be
F_w^{-1}=\frac{(1+w)^2}{\tilde{C}(1+2w)(1+w+w^2)}\left[I\otimes I-\frac{w}{1+w} V\otimes U +\frac{w^2}{(1+w)^2} U\otimes V\;\right].
\ee
The twist $F_w$ satisfies the YBE, guaranteeing that the $R$-matrix is quasitriangular (actually
triangular, as discussed in the above) and that the
twisted coproduct $\Delta_F$ will be associative. We also note that it satisfies both relations
in (\ref{RFrels}).

Although the twist as given in (\ref{wtwist}) is perfectly acceptable,  in order to better exhibit its
properties and compare to other twists in the literature, it is useful to write it in exponential form. 
Using the special properties of the shift matrices, in particular the fact that 
$(V\otimes U)^2=U\otimes V$ and $(V\otimes U)^3=I\otimes I$, we can rewrite $F_w$ as\footnote{We stick to
the $U$ and $V$ notation here for simplicity, but of course when writing $F_w$ in exponential form it 
would also be appropriate to use the hermitian algebra generators $X=U+V$ and $Y=i(U-V)$ instead.} 
\be \label{Fwexponential}
F_w= e^{a(w) V\otimes U + b(w) U\otimes V}\;,
\ee
where
\be
a(w)=\frac{1}{6}\ln \left(\frac{(1+2w)^2}{1+w+w^2}\right)+{{\arctan \left({{1+2w
 }\over{\sqrt{3}}}\right)}\over{\sqrt{3}}}-\frac{\pi}{6\sqrt{3}}
\ee
and 
\be
b(w)=\frac{1}{6}\ln \left(\frac{(1+2w)^2}{1+w+w^2}\right)-{{\arctan \left({{1+2w
 }\over{\sqrt{3}}}\right)}\over{\sqrt{3}}}+\frac{\pi}{6\sqrt{3}}\;.
\ee
In order to be able to write the twist in this form, i.e. as an exponential without a prefactor, it
is necessary to choose the normalisation constant appropriately. The requisite choice is
\be
\tilde{C}=\frac{C(a(w))}{1+w}
\ee
where $C(x)$ is the hypergoniometric cosine defined in appendix \ref{Hypergoniometric}. We will also
use the corresponding hypergoniometric sine $S(x)$, which satisfies $S'(x)=C(x)$ as well
as 
\be
S(x)^3+C(x)^3=1
\ee
which is analogous to the usual trigonometric identity. Substituting the expression for $a(w)$ above,
we find
\be \label{CSaw}
C(a(w))=\frac{1+w}{[(1+2w)(1+w+w^2)]^{\frac13}}\;,\quad S(a(w))=\frac{w}{[(1+2w)(1+w+w^2)]^{\frac13}}\;.
\ee
We could thus simply have defined $\tilde{C}=[(1+2w)(1+w+w^2)]^{-1/3}$, but we find that expressing the
twist in terms of $C(a)$ and $S(a)$ is more appealing, in that it highlights the nilpotent nature of our shift
matrices $U$ and $V$ (see appendix \ref{Hypergoniometric}). 

After these preliminaries, it can be checked that expanding the exponential form of $F_w$ in 
terms of $a$ and $b$ the terms proportional to $U\otimes V$ cancel out and the series organises itself as 
\be
F=C(a) I\otimes I +S(a) V\otimes U=C(a)\left[I\otimes I+\frac{S(a)}{C(a)} V\otimes U\right]\;.
\ee
Inserting the dependence of $a$ on $w$ above we find that (see (\ref{CSaw}))
\be
\frac{S(a)}{C(a)}=\frac{w}{1+w}
\ee
which gives us back the original twist (\ref{wtwist}) and shows its equivalence to
(\ref{Fwexponential}). As before, writing $F_w$ in the form (\ref{Fwexponential}) also
guarantees that it is counital, i.e. that $(\epsilon\otimes \text{id})F_w=1$, since $\epsilon$
exponentiates and  $\epsilon(U)=\epsilon(\half(X-iY))=0$, $\epsilon(V)=\epsilon(\half(X+iY))=0$. So $F_w$ is indeed a Hopf twist.

It is also straightforward to check that the $R$-matrix can be expressed as
\be
R_w=F_{w,21} F^{-1}_{w,12}=e^{(a(w)-b(w))U\otimes V+(b(w)-a(w))V\otimes U}\;,
\ee
which Taylor expands to give precisely (\ref{Rw}).

If we now define the twisted coproduct as usual to be given by
\be
\Delta_F^{(w)}=F_w \Delta F_w^{-1}
\ee
we find that on functions of $U$ and $V$ the coproduct is the same as the undeformed one,
again  due to the abelian nature of the matrices involved. So the twisted coproduct
is trivially associative.

Finally, from the above we can  easily show that:
\be
(\Delta_F\otimes\mathrm{id}) R_w=R_{w,13}R_{w,23}\;,\quad (\mathrm{id}\otimes \Delta_F)R_w=R_{w,13}R_{w,12}
\ee
which is the quasitriangularity condition. Of course, this condition implies the YBE for $R_w$, which
we already know to be satisfied, but we believe that the more detailed understanding of the
Hopf algebra structure presented in the above will turn out to be useful in extending our results to other cases.  

The theoretical framework, mainly due to Drinfeld,
 of how quantum groups arise from twists of other quantum groups and in particular Lie algebras,
is well understood. However,  there are very few \emph{explicit} constructions in the literature. 
Even the twist from $U(\su(2))$ (the universal enveloping algebra of $\su(2)$) to $U_q(\su(2))$ is not available 
in full. The work \cite{Dabrowskietal96} provides such a twist  to second order,
though it does not satisfy the cocycle condition. In practice, one is typically interested in
specific matrix representations of the twist, and in that case there do exist explicit results, 
such as \cite{Engeldinger95} for the tensor product of the fundamental representation and 
\cite{Chryssomalakos97} for the tensor product of the adjoint representation (for $U_q(\su(2))$). 
Drinfeld twistings of $U(sl(3))$ are classified in \cite{Kulishetal06}. 
The twists we have discussed explicitly deform $U(\su(3))$ to the (equivalent) Hopf algebras $U_q(\su(3))$ and 
$U_w(\su(3))$, in the tensor product of the fundamental. 

Having  concluded our construction of the $w$-twist, let us now discuss how twists can be used
to define a non-commutative star product on the coordinates of $\Cset_3$.

\section{The star product from twisting} \label{StarProduct}

In general, the twisting procedure outlined above leads to a new, deformed Hopf (or quasi-Hopf) algebra. In 
the specific case we are considering, the undeformed Hopf algebra is simply the Lie group $\SU(3)$, the symmetry 
group acting on the three $\Ncal=4$ SYM scalars $\Phi^i$. This Hopf algebra clearly has a trivial coproduct, acting as 
$\Delta(X)=X\otimes I+I\otimes X$ on the $\su(3)$ algebra elements. Twisting leads to a non-trivial Hopf
algebra, which was studied in some detail in \cite{Mansson:2008xv}. In particular, the commutation relations
of the algebra were constructed from the twisted $(q,h)$ $R$-matrix using the RTT relations \cite{FRT90}, and it was
shown that the quantum determinant is a central element, from which it follows that the Hopf algebra is a
global symmetry of the deformed superpotential.\footnote{The work \cite{Mansson:2008xv} worked
in the dual picture to that employed here, where the noncommutative coproduct is matrix multiplication, while
it is the \emph{product} between matrix elements which is commutative in the undeformed case and becomes non-commutative
after twisting. The two pictures are of course equivalent in our bialgebra setting.} 

 In the following we will not be interested in the Hopf algebra itself, but rather 
in the function space on which it acts (i.e. the algebra module). This space is spanned by three 
complex coordinates $z^i$. Before twisting, these are just coordinates on $\Cset_3$, in line with 
the symmetry algebra being $\su(3)$. After the twist, they become non-commutative coordinates of a quantum plane, 
with its symmetry algebra being the Hopf algebra $\widetilde{\SU(3)}_{q,h}$ \cite{Mansson:2008xv}. To better study this
non-commutative structure, we would now like to find a star product among the $z^i$ coordinates which
reproduces the quantum plane relations derived in \cite{Mansson:2008xv}. This will be the standard star product
arising in deformation quantisation, which involves the twist in a fundamental way. Since so far
we only have an explicit construction of the twist for the integrable $\beta$- and $w$-deformations, in
what follows we will restrict our attention to those cases. The following construction follows
\cite{Majid,Blohmann03}, which can be consulted for further details.  

The notion that we want to preserve while twisting is that of compatibility between the coproduct of
the Hopf algebra $\mathcal{H}$ and the product of the module algebra $A$. The elements of the Hopf algebra act on the 
module in the usual way as derivations, i.e. for $g\in \mathcal{H}$ and $x,y\in A$ we have
\be
g\triangleright (x\cdot y)= (g\triangleright x) \cdot y + x\cdot (g\triangleright y)\;.
\ee
Where we used the undeformed coproduct $\Delta(g)=g\otimes 1+1\otimes g$. Twisting 
the coproduct as in (\ref{DeltaF}) naturally induces a compatible twisting of the \emph{product} on
the algebra module. To see this, consider the above action in slightly more detail:\footnote{We 
will use the notation $X\cdot Y$ and $m(X\otimes Y)$ interchangeably to denote module multiplication.}
\be
g\triangleright (x\cdot y)= m(\Delta(g) \triangleright [x\otimes y])=m( [g\otimes 1 +1\otimes g]\triangleright [x\otimes y])
=m( (g\triangleright x)\otimes y + x\otimes (g\triangleright y))= (g\triangleright x)\cdot y+x\cdot (g\triangleright y)
\ee
Since we want the product rule property to hold when the coproduct is twisted as $\Delta\ra F \Delta F^{-1}$,
 we find that we also need to twist the module product as \cite{Majid,Blohmann03}:\footnote{ 
See also \cite{Watts:1999ra,Watts:2000mq} for a similar construction. Although the focus there is
 on the action of the Hopf algebra on itself rather than the module, we have found the exposition in those
works
instructive.}
\be
m_F (x\otimes y) =m( F^{-1} \triangleright x\otimes y)=(F^{-1}_{(1)}\triangleright x)\cdot (F^{-1}_{(2)}\triangleright y)
\ee
To show this, one can start by assuming the product rule property to obtain:
\be
\begin{split}
g\triangleright m_F(x\otimes y)&=m(F^{-1} F\Delta(g)F^{-1}\triangleright [x\otimes y])=
m\left([gF^{-1}_{(1)}\otimes F^{-1}_{(2)}+F^{-1}_{(1)}\otimes g F^{-1}_{(2)}] \triangleright [x\otimes y]\right)\\
&=(gF^{-1}_{(1)}\triangleright x)\cdot (F^{-1}_{(2)}\triangleright y)+(F^{-1}_{(1)}\triangleright x)\cdot (gF^{-1}_{(2)}\triangleright y)
\end{split}
\ee
Alternatively, one can perform the same computation by first expanding out the twisted product:
\be
\begin{split}
g\triangleright m_F(x\otimes y)&=g\triangleright m(F^{-1}_{(1)}\triangleright x\otimes F^{-1}_{(2)}\triangleright y)
=m(\Delta(g)\triangleright [F^{-1}_{(1)}\triangleright x\otimes F^{-1}_{(2)}\triangleright y])\\
&=m(g F^{-1}_{(1)} \triangleright x\otimes F^{-1}_{(2)}\triangleright y
+F^{-1}_{(1)} \triangleright x \otimes g F^{-1}_{(2)}\triangleright y)
\end{split}
\ee
which leads to the same answer. This confirms the compatibility of the twisted Hopf algebra 
coproduct with the twisted module product.

We then define our star product on the module to be precisely the twisted product above, i.e. 
\be \label{starproduct}
x\star y =m_F(x\otimes y)\;.
\ee
For our purposes, it will often be sufficient to consider the action of the twist on the module
coordinates $z^i$, rather than on arbitrary functions. 
In that case, it can be convenient to re-express this relation in index notation:
\be \label{starproductindex}
z^i\star z^j=(F^{-1})^{j\;i}_{\;l\;\;k} z^l z^k
\ee
Note that here the coordinates on the right-hand side are commuting, so the ordering of the $k,l$ indices is unimportant.

\subsection{Compatibility with the quantum plane relation} \label{compatibility}

Let us check that this definition of the star product is consistent with the quantum plane structure. Recall
that in (\ref{qplane}) the $z^i$ are quantum plane coordinates, which are non-commuting. For many physical
applications (including the application we have in mind) it is very desirable to 
transfer the noncommutativity from the coordinates to the star product, i.e. we would like the star
product to lead to the following relation:
\be \label{qplanestar}
z^j\star z^i=R^{i\;j}_{\;k\;\;l}z^k\star z^l\;.
\ee
Here commutative $z^i$'s are multiplied with a non-commutative product. It is easiest to show that our
star product has this property using its index form (\ref{starproductindex}). We have:
\be \label{compatibilityeq}
z^j\star z^i=(F^{-1})^{i\;j}_{\;m\;n} z^m z^n= (F^{-1})^{i\;j}_{\;m\;n} z^n z^m= (F^{-1})^{i\;j}_{\;m\;n} (F^{nm}_{\;l\;\;k} z^k\star z^l)
=R^{i\;j}_{\;k\;\;l} z^k\star z^l
\ee
where we used the inverse of (\ref{starproductindex}) as well as the second relation in (\ref{RFrels}),
which can be seen to be of crucial importance for our construction. As both our $\beta$ and $w$-twists
satisfy (\ref{RFrels}), we expect compatibility between the star product and the quantum plane for these
cases. 

\subsection{Higher-order star products}

Of course, we are also interested in star products between more than two coordinates. This is where
the potential for ambiguities due to non-associativity arises, since different placements of parentheses
can lead to different answers. However, as discussed in appendix \ref{HigherStarProducts}, the cocycle identity 
guarantees associativity. Let us apply the results of that appendix to the star product between three 
coordinates, choosing for concreteness the case 
where they are all different. We find that the two possible orderings of parentheses are equal to 
\be
\begin{split}
(z^1\star z^2)\star z^3&=(F^{-1})^{32}_{\;k'j'} (F^{-1})^{k'1}_{\;ni'}(F^{-1})^{j'i'}_{ml}z^{l}z^{m}z^{n}\\
z^1\star (z^2\star z^3)&=(F^{-1})^{21}_{\;j'i'}(F^{-1})^{3i'}_{k'l} (F^{-1})^{k'j'}_{nm} z^{l}z^{m}z^{n}\;.
\end{split}
\ee
If our  twist satisfies the YBE, the two placements are equal and there is a unique star product between 
three coordinates. We can then pick one ordering and remove the parentheses to write just
\be \label{cubicstar}
z^1\star z^2\star z^3=(F^{-1})^{21}_{\;j'i'}(F^{-1})^{3i'}_{k'l} (F^{-1})^{k'j'}_{nm} z^{l}z^{m}z^{n}\;.
\ee
We can similarly derive expressions for the higher-order star products acting on coordinates. (See
appendix \ref{HigherStarProducts} for some of these expressions). The number of twist matrices in a 
star product between $d$ coordinates is ${\tiny\doublet{d}{2}} =\frac{d!}{2!(d-2)!}$,
as can be expected by expanding the twists to first order, in which case the star product reduces
to all possible single Wick contractions between the coordinates. 
 The ordering of the twist matrices will depend on the placement of the parentheses, however the YBE relations 
between the twists guarantee that all orderings are equal, and thus that the star product is associative. 

In defining the above star product we have followed a standard construction in deformation quantisation.
However, since our focus has been on the action of the star product on the coordinates $z^i$, and 
not on general functions, it does not take the differential form derived by Kontsevich \cite{Kontsevich:1997vb}. 
Leaving the extension to functions and comparison with \cite{Kontsevich:1997vb} 
to future work, we will now turn to explicit expressions of the star
product in the two specific cases under consideration.

\subsection{The $\beta$-twisted star product}

For the diagonal $\beta$-twist we will use the explicit index notation (\ref{starproductindex}). We
obtain
\be
z^1\star z^2 =(F^{-1})^{2\;1}_{\;2\;1} z^2 z^1=q^\half z^1 z^2 \;\quad 
z^2\star z^1=(F^{-1})^{1\;2}_{\;1\;2} z^1 z^2=q^{-\half} z^1 z^2
\ee
The remaining relations follow by cyclicity. Combining the two relations, we find 
\be
z^2\star z^1=q^{-1} z^1\star z^2 = R_q{}^{1\;2}_{\;1\;2} z^1 \star z^2
\ee
in agreement with the quantum plane relation (\ref{qplanestar}). We can also compute
\be
z^1\star z^2\star z^3=(F^{-1})^{2\;1}_{\;2\;1}(F^{-1})^{3\;1}_{\;3\;1}(F^{-1})^{3\;2}_{\;3\;2} z^1z^2 z^3=q^{\half} z^1 z^2 z^3
\ee
and
\be
z^1\star z^3\star z^2=(F^{-1})^{3\;1}_{\;3\;1}(F^{-1})^{2\;3}_{\;2\;3}(F^{-1})^{2\;1}_{\;2\;1} z^1z^3 z^2=q^{-\half} z^1 z^2 z^3
\ee
for the two cyclically distinct combinations. Switching to gauge theory notation for purposes of
illustration, we can use these results to show that the $q$-deformed superpotential with the fields
multiplied with star products is equal (up to an unimportant factor)
to the undeformed $\Ncal=4$ superpotential with the usual product between fields:
\be
\Tr\left(\Phi^1\star \Phi^2\star \Phi^3-q \Phi^1\star \Phi^3\star \Phi^2\right)=
q^\half \Tr\left(\Phi^1[\Phi^2,\Phi^3]\right)
\ee

\subsection{The $w$-twisted star product}

For the $w$-twist, the explicit form of the star product is:
\be
x\star y= m(F_w^{-1}\triangleright [x\otimes y])=m(x\otimes y) -\frac{w}{1+w} m( V~ x\otimes U~y)+\frac{w^2}{(1+w)^2}
m(U~x\otimes V~y) 
\ee
Now let us specialise to the case of the star product between the coordinates $z^i$ spanning the module. So
from now on we focus on the fundamental representation.  
To consider the action of the matrices $U$ and $V$ on the module, we use the basis
\be
z^1=\triplet{1}{0}{0}\;,\quad z^2=\triplet{0}{1}{0}\;,\quad z^3=\triplet{0}{0}{1}
\ee
so that e.g.\footnote{Equivalently, one can express the action of $U$ and $V$ on the $z^i$ by
re-writing them as differential operators: $
U=z^1\p_3+z^2\p_1+z^3\p_2 \;,\quad \; V=z^1\p_2+z^2\p_3+z^3\p_1$.}
\be
Vz^1=\threebythree{0}{1}{0}{0}{0}{1}{1}{0}{0} \triplet{1}{0}{0}=\triplet{0}{0}{1}=z^3
\ee
We thus find:
\be
\begin{split}
z^1\star z^2&=z^1z^2-\frac{w}{1+w} (z^3)^2+\frac{w^2}{(1+w)^2} z^1z^2=\left(1+\frac{w^2}{(1+w)^2}\right)z^1z^2-\frac{w}{1+w}(z^3)^2\\
z^2\star z^1&=z^1z^2-\frac{w}{1+w} z^1z^2+\frac{w^2}{(1+w)^2}(z^3)^2=\left(1-\frac{w}{1+w}\right)z^1z^2 +\frac{w^2}{(1+w)^2}(z^3)^2\\
z^3\star z^3&=(z^3)^2-\frac{w}{1+w} z^1z^2=\frac{w^2}{(1+w)^2}z^1z^2=-\frac{w}{1+w}\left(1-\frac{w}{1+w}\right)z^1z^2+(z^3)^2
\end{split}
\ee
Here we have suppressed the overall normalisation of each expression, since we are interested in writing relations purely between the star products and the normalisation drops out. Eliminating $(z^3)^2$ and $z^1z^2$ 
from the above expressions we find:
\be
z^1\star z^2=(1+w)~z^2\star z^1-w~ z^3\star z^3
\ee
As expected, we have found the star-product version (\ref{qplanestar}) of the quantum plane relation (\ref{qplane}). 

Let us also look at the star product (\ref{cubicstar}) 
of three coordinates, for which we need to compute the explicit expression:  
\be
\begin{split}
F^{-1}_{12}F^{-1}_{13}F^{-1}_{23}=&\frac{1+w}{(1+2w)^2(1+w+w^2)}\big[(1+4 w+5 w^2+ w^3) I \otimes I \otimes I \\
&   + w^2(1+w) (  I \otimes U \otimes V+U \otimes V \otimes I + U \otimes I \otimes V+ U \otimes U \otimes U+ V \otimes V \otimes V)   \\
&-w(1+3w+3w^2) (I \otimes V \otimes U+V \otimes U \otimes I) 
-  \frac{w(1+3w+2w^2-w^3)}{1+w} V \otimes I \otimes U  \big]
\end{split}
\ee
Let us apply this to the $w$-deformed superpotential, where again we temporarily 
switch to gauge theoretical language by replacing the $z^i$ coordinates with the $\Phi^i$ superfields. A 
straightforward computation gives:
\be
\begin{split}
&\Phi^1\star\Phi^2\star\Phi^3\!+\!\Phi^2\star\Phi^3\star\Phi^1\!+\!\Phi^3\star\Phi^1\star\Phi^2
-(1\!+\!w)\left(\Phi^1\star\Phi^3\star\Phi^2\!+\!\Phi^3\star\Phi^2\star\Phi^1\!+\!\Phi^2\star\Phi^1\star\Phi^3\right)\\
&\quad+w\left(\Phi^1\star\Phi^1\star\Phi^1+\Phi^2\star\Phi^2\star\Phi^2+\Phi^3\star\Phi^3\star\Phi^3\right)\\
&=(1+2w)\left[\Phi^1\Phi^2\Phi^3+\Phi^2\Phi^3\Phi^1+\Phi^3\Phi^1\Phi^2-\Phi^1\Phi^3\Phi^2-\Phi^3\Phi^2\Phi^1-\Phi^2\Phi^1\Phi^3\right]
\end{split}
\ee
Inserting the above expression into the gauge-theory trace, we see the equivalence (up to overall normalisation) 
of the $w$-deformed superpotential,
where the matrices are multiplied with the star product, with the undeformed $\Ncal=4$ superpotential. 
In a way the role of the star product is to show that the deformed theory is equivalent to the undeformed
one, as long as we are prepared to multiply the fields of the deformed theory with this noncommutative product.

\subsection{The inverse star product}

We saw above how the star product (\ref{starproduct}), which is standard in the context of deformation quantisation,
expresses the non-commutativity of the quantum plane coordinates. Deformed expressions (such as the
quantum plane relations or the deformed superpotential) reduce to the undeformed ones when the coordinates
are multiplied with this star product. In this section, we will consider an inverse question: Can we take
an undeformed expression and, by substituting the product with an appropriate star-product, convert it to a 
(meaningful) deformed expression? Having this ability will be very useful if we want to use the star product as a
solution-generating technique, since we would like to take an undeformed ($\Ncal=4$ SYM in our specific context)
expression and obtain a deformed one (relevant to the $\beta$ or $w$-deformation in our context) simply by
inserting a star product. 

More precisely, what we require is a star product which, inserted into a valid undeformed expression, leads to 
a valid deformed expression in which there is no star product but the coordinates are themselves noncommutative.
This is the inverse procedure to that discussed in section \ref{compatibility}.

 The star product that achieves this is simply the inverse star product to the one above. We define it as
\be \label{asteriskproduct}
x \ast y=m( F \triangleright x\otimes y) \qquad \left(\text{in indices:} \quad z^i\ast z^j=F^{ji}_{\;lk} z^k z^l\right)
\ee
where we denote the inverse star product with an asterisk instead of a star. 

Let us check that taking an obvious undeformed expression and inserting this star product 
leads to the corresponding quantum plane:
\be
z^iz^j=z^j z^i \;\ra\; z^i\ast z^j=z^j\ast z^i\; \Rightarrow\; F^{j\;i}_{\;l\;k} z^k z^l=F^{i\;j}_{\;m\;n} z^n z^m
\ee
where in the last expression the $z^i$ are noncommutative. Multiplying by the inverse of $F$, we find
\be
z^n z^m=(F^{-1})^{m\;n}_{\;i\;j} F^{j\;i}_{\;l\;k} z^k z^l=R^{m\;n}_{\;k\;l} z^k z^l
\ee
which is the required quantum plane relation. Using the $\ast$ product, we have thus performed 
the precisely opposite procedure to (\ref{compatibilityeq}). 

Repeating the procedure discussed above for the cubic terms, we find that 
\be
z^1\ast z^2\ast z^3=m\left(F_{12}F_{13}F_{23}\triangleright [z^1\otimes z^2\otimes z^3]\right)
\ee
where for the $w$-deformation
\be
\begin{split}
F_{12}F_{13}F_{23}&=\tilde{C}^3(1+w)^3\big[
I\otimes I\otimes I+\frac{w}{1+w} (I\otimes V\otimes U+V\otimes U\otimes I)
+\frac{w^3}{(1+w)^3} U\otimes I\otimes V\\
&+\frac{w^2}{(1+w)^2} (U\otimes U\otimes U+V\otimes V\otimes V)
+\frac{w(1+2w)}{(1+w)^2} V\otimes I\otimes U
    \big]\;.
\end{split}
\ee
We can now straightforwardly check that the 
$\Ncal=4$ SYM superpotential deforms (up to overall normalisation) to the $w$-deformed one
simply by converting all the normal products to inverse star products:
\be \label{fromundeformedtodeformed}
\begin{split}
&\Phi^1\ast\Phi^2\ast\Phi^3+\Phi^2\ast\Phi^3\ast\Phi^1+\Phi^3\ast\Phi^1\ast\Phi^2-
\Phi^1\ast\Phi^3\ast\Phi^2-\Phi^3\ast\Phi^2\ast\Phi^1-\Phi^2\ast\Phi^1\ast\Phi^3\\
&=\frac{1}{1+2w}\left[\Phi^1 \Phi^2 \Phi^3+\Phi^2 \Phi^3 \Phi^1+\Phi^3 \Phi^1 \Phi^2-
(1+w)\left[\Phi^1 \Phi^3 \Phi^2-\Phi^3 \Phi^2 \Phi^1-\Phi^2 \Phi^1 \Phi^3\right]\right.\\
&\left.\qquad\qquad\qquad\quad+w\left((\Phi^1)^3+(\Phi^2)^3+(\Phi^3)^3\right)\right]
\end{split}
\ee
Taking the gauge theory trace on both sides leads to the expression
\be
\begin{split}
\Tr &\left(\Phi^1\ast\Phi^2\ast  \Phi^3-\Phi^1\ast \Phi^3\ast \Phi^2\right)\\
&=\frac{3}{1+2w}\Tr \left(\Phi^1 \Phi^2  \Phi^3-(1+w)\Phi^1 \Phi^3 \Phi^2
+\frac{w}{3}\left((\Phi^1)^3+(\Phi^2)^3+(\Phi^3)^3\right)\right)
\end{split}
\ee
where the left-hand side is schematic and denotes the trace of the left-hand side of (\ref{fromundeformedtodeformed}).

It is of course very appealing to be able to write a Leigh-Strassler-deformed superpotential
as simply the $\Ncal=4$ SYM superpotential, just with the products between fields replaced by star products.  
Let us remark that a star product achieving the same result for the integrable deformations has previously been proposed 
in \cite{Bundzik:2006jz}. It would be interesting to explore the relation between these products and, in particular,
whether the product in \cite{Bundzik:2006jz} can be derived from the perspective of a twist.    

In the following we will use the star product (\ref{asteriskproduct}) to deform the pure spinors of the undeformed background to 
those of the deformed one.

\section{The first-order star product} \label{Noncommutativity}

 After our study of the all-orders twist, in this section we will expand it to first order, so as to make 
contact with previous work on non-commutativity based on the classical $R$-matrix of the theory.
We will focus exclusively on the $\ast$-product defined in (\ref{asteriskproduct}), since this is
the product which, inserted into an undeformed expression, leads to its appropriately deformed version. However,
we note that at first order the $\ast$ and $\star$ products only differ by a sign. 

\subsection{The $\beta$-deformation}

Although the first-order structure of the $\beta$-deformation is standard, we will review it here
for completeness. We start by expanding the twist (\ref{qTwist})
\be
F\mbox{}^{i\;\;j}_{\;k\;\;l}=\delta^i_{\;\;k}\delta^j_{\;\;l}-\frac{i\beta}{2}~r^{i\;\;j}_{\;k\;\;l}
\ee
with $r$ the classical $r$-matrix, $r=\text{diag}(0,-1,1,1,0,-1,-1,1,0)$. The first-order
star product is then
\be
z^i\ast z^j=z^i z^j -\frac{i\beta}2 r^{j\;i}_{\;l\;k} z^k z^l=z^i z^j+\frac{i\beta}2 r^{i\;j}_{\;k\;l}z^k z^l
\ee
where we have used the fact that $r_{21}=-r_{12}$ for the classical $r$-matrix. We can now express
the commutator of two coordinates as
\be
z^i\ast z^j-z^j\ast z^i=\Theta^{ij}_\beta\;, 
\ee
where $\Theta^{ij}_\beta=i\beta r^{i\;\;j}_{\;k\;l}z^k z^l$ is the \emph{non-commutativity matrix} \cite{Kulaxizi:2006pp}:
\be \label{betaThetamatrixhol}
\Theta^{ij}=i\beta\threebythree{0}{-z^1z^2}{z^1 z^3}{z^1 z^2}{0}{-z^2 z^3}{-z^1z^3}{z^2z^3}{0}
\ee
Note that, despite the resemblance to more familiar star products, here $\Theta^{ij}$ is a function of the coordinates. We
will see that, after extending it to the antiholomorphic sector, it will transform to a constant matrix in an 
appropriate coordinate system.

\subsection{The $w$-deformation} \label{wdeffirstorder}

Expanding the $w$-twisted $R$-matrix (\ref{wRmatrix})  to first order gives us
\be
F\mbox{}^{i\;\;j}_{\;k\;\;l}=\delta^i_{\;\;k}\delta^j_{\;\;l}+w~r^{i\;\;j}_{\;k\;\;l}
\ee
where the classical $r$-matrix is
\be
r=\begin{pmatrix}0&0&0&0&0&-1&0&1&0\cr 0&0&0&-1&0&0&0&0&1\cr 0&0&0&0&-1&0&1
 &0&0\cr 0&1&0&0&0&0&0&0&-1\cr 0&0&1&0&0&0&-1&0&0\cr 1&0&0&0&0&0&0&-1
 &0\cr 0&0&-1&0&1&0&0&0&0\cr -1&0&0&0&0&1&0&0&0\cr 0&-1&0&1&0&0&0&0&0
 \cr \end{pmatrix}
\ee
and can of course also be obtained from the $\beta$-twisted $r$ matrix above by a similarity transformation with $T\otimes T$
(and a rescaling of $\sqrt{3}$).  
It is easy to see that $r_{12}=f_{21}-f_{12}$, where $f$ are the first-order expansions of the corresponding twists.
So we find:
\be
z^i\ast z^j-z^j\ast z^i=w(f^{j\;i}_{\;l\;\;k}-f^{i\;j}_{\;k\;\;l})z^kz^l= w r^{i\;j}_{\;k\;\;l} z^kz^l\;.
\ee
Defining  again our (holomorphic) noncommutativity matrix as $\Theta^{ij}=w r^{i\;\;j}_{\;k\;\;l} z^k z^l$, we obtain
\be \label{Thetawholomorphic}
\Theta^{ij}_w=w\begin{pmatrix}0&(z^3)^2-z^1\,z^2&z^1\,z^3-(z^2)^2\cr 
z^1\,z^2-(z^3)^2&0&(z^1)^2-z^2\,z^3\cr (z^2)^2-z^1\,z^3&z^2\,z^3-(z^1)^2&0\cr\end{pmatrix}
\ee
with which we write the first-order noncommutativity relation as:
\be
z^i\ast z^j-z^j\ast z^i=\Theta^{ij}_w\;.
\ee
Note the absence of an $i$ in $\Theta^{ij}_w$ as compared to $\Theta^{ij}_\beta$. This can be traced to the reality
of $q$ and $h$ in the $w$ deformation, while for the $\beta$-deformation $q$ is a phase and expands as $i \beta$. 

\subsection{Mixed coordinates}

  In the above, we only considered the twisting relevant to the holomorphic coordinates, which in gauge theory 
language corresponds to twisting the superpotential, which is a function of the holomorphic superfields $\Phi^i$. 
However, since the full gauge theory potential also depends on the antiholomorphic fields $\bar{\Phi}^i$, 
to get the full picture we need to consider functions of the antiholomorphic coordinates as well. As discussed in 
\cite{Mansson:2008xv}, their noncommutativity 
properties can be simply derived from the definitions of the antiholomorphic and mixed quantum planes respectively. 
Denoting (only in this section) the antiholomorphic coordinates $\bar{z}^{\bar{i}}$ with an index down, 
$\bar{z}_i=\bar{z}^{\bar{i}}$, we have
\be \label{Ruu}
\zb_k \zb_l \R kilj=\zb_j \zb_i
\ee
for the antiholomorphic quantum plane, and\footnote{Here $\widetilde{R}$ is the \emph{second inverse} of $R$, 
defined as $
\Rt imnj \R mlkn=\delta^i_{\;l}\delta^k_{\;j}=\R imnj \Rt mlkn\;$.} 
\be \label{MixedPlane}
\zb_l \R jkli z^k=z^j \zb_i\; \quad \text{and}\quad z^k \Rt iklj \zb_l=\zb_j z^i\;.
\ee
for the mixed quantum planes. As discussed in more detail in \cite{Mansson:2008xv}, the mixed plane 
relations are consistent with each other and are also invariant under $\widetilde{\SU(3)}_w$ transformations. 
The gauge theory relevance of the mixed relations is that although the holomorphic and antiholomorphic star products
are designed to deform the $F$-terms of the theory, when we convert all the products in the potential to star products
they also end up deforming terms
that should not be affected by the Leigh-Strassler deformation (in particular, D-terms). The mixed terms cancel 
those unwanted contributions, ensuring that the end result of the deformation is a Leigh-Strassler theory. 

Expanding the $R$ matrices in the above relations to first order and taking commutators, we are led to the
following set of noncommutativity relations:
\be
[z^i, z^j]_\ast=\Theta^{ij}\;,\quad [z^i, \zb^{\bar{j}}]_\ast=\Theta^{i\bar{j}}\;,\quad 
[\zb^{\bar{i}}, z^j]_\ast=\Theta^{\bar{i}j}\;,\quad
[\zb^{\bar{i}}, \zb^{\bar{j}}]_\ast=\Theta^{\bar{i}\bar{j}}\;.
\ee
We can combine all these relations into just one as
\be \label{Thetastar}
[z^I,z^J]_\ast=\Theta^{IJ}
\ee
where the indices $I,J$ range from $1,2,3,\bar{1},\bar{2},\bar{3}$ and  $z^{\bar{i}}=\zb^i$. 

\mparagraph{The $\beta$-deformation}

For the (real) $\beta$-deformation the full noncommutativity matrix was derived in \cite{Kulaxizi:2006pp}
and reads:
\be \label{betaTheta}
\Theta^{IJ}_\beta=i \beta \begin{pmatrix}0&-{  z^1}\,{  z^2}&{  z^1}\,{  z^3}&0&{  z^1}\,
 \zb^2&-{  z^1}\,\zb^3\cr {  z^1}\,{  z^2}&0&-{  z^2}
 \,{  z^3}&-\zb^1\,{  z^2}&0&{  z^2}\,\zb^3\cr -
 {  z^1}\,{  z^3}&{  z^2}\,{  z^3}&0&\zb^1\,{  z^3}&-
 \zb^2\,{  z^3}&0\cr 0&\zb^1\,{  z^2}&-\zb^1\,
 {  z^3}&0&-\zb^1\,\zb^2&\zb^1\,\zb^3\cr -{  z^1}
 \,\zb^2&0&\zb^2\,{  z^3}&\zb^1\,\zb^2&0&-\zb^2
 \,\zb^3\cr {  z^1}\,\zb^3&-{  z^2}\,\zb^3&0&-
 \zb^1\,\zb^3&\zb^2\,\zb^3&0\cr \end{pmatrix}
\ee
In \cite{Mansson:2008xv} it was checked that the mixed and antiholomorphic plane relations also produce this matrix 
as the unique extension of (\ref{betaThetamatrixhol}).
As shown in \cite{Kulaxizi:2006pp}, switching to a spherical coordinate basis on $\Cset_3$ as
\be \label{sphericalcoords}
z^1=r \cos\alpha e^{i\phi_1}\;,\quad z^2=r\sin\alpha\sin\theta e^{i\phi_2}\;,\quad z^3=r \sin\alpha\cos\theta e^{i\phi_3}
\ee
converts $\Theta^{IJ}_\beta$ to a coordinate-independent matrix, which furthermore acts purely on the the three angles $\phi^i$:
\be
\Theta^{IJ}_\beta \frac{\p}{\p z^I}\wedge \frac{\p}{\p z^J}=\beta^{ij}\frac{\p}{\p \phi^i}\wedge\frac{\p}{\p \phi^j}\;,
\quad \text{with} \quad \beta^{ij}=\threebythree{0}{\beta}{-\beta}{-\beta}{0}{\beta}{\beta}{-\beta}{0}\;.
\ee
This shows that for the $\beta$-deformation the star product (\ref{Thetastar}) is nothing but
the Lunin-Maldacena star product \cite{Lunin:2005jy}, and also explains why there is no issue with
non-associativity for real $\beta$: In an appropriate coordinate system,
the matrix is simply constant.

\mparagraph{The $w$-deformation}

For the $w$-deformation, we find the full noncommutativity matrix, which extends (\ref{Thetawholomorphic}) by including
the mixed and antiholomorphic planes, to be: 
\be \label{wThetamatrix}
\Theta^{IJ}_w=w\begin{pmatrix}0&(z^3)^2-z^1\,z^2&z^1\,z^3-(z^2)^2&z^2\,\bar{z}
 ^2-z^3\,\bar{z}^3&z^2\,\bar{z}^3-\bar{z}^1\,z^3&
 \bar{z}^1\,z^2-\bar{z}^2\,z^3\cr z^1\,z^2-(z^3)^2&0&(z^1)^2-z^2\,z^3&\bar{z}^2\,z^3-z^1\,\bar{z}^3&z^3\,
 \bar{z}^3-z^1\,\bar{z}^1&\bar{z}^1\,z^3-z^1\,\bar{z}^2
 \cr (z^2)^2-z^1\,z^3&z^2\,z^3-(z^1)^2&0&z^1\,\bar{z}^2-z^2\,\bar{z}^3&z^1\,\bar{z}^3-\bar{z}^1\,z^2&z^1\,
 \bar{z}^1-z^2\,\bar{z}^2\cr z^3\,\bar{z}^3-z^2\,\bar{z}^2&z^1\,\bar{z}^3-\bar{z}^2\,z^3&z^2\,\bar{z}^3-z^1
 \,\bar{z}^2&0&\bar{z}^1\,\bar{z}^2-(\bar{z}^3)^2&(\bar{z}^2)^2
 -\bar{z}^1\,\bar{z}^3\cr 
\bar{z}^1\,z^3-z^2\,\bar{z}^3&z^1\,\bar{z}^1-z^3\,\bar{z}^3&\bar{z}^1\,z^2-z^1\, \bar{z}^3
&(\bar{z}^3)^2-\bar{z}^1\,\bar{z}^2&0&\bar{z}^2\,
 \bar{z}^3-(\bar{z}^1)^2\cr \bar{z}^2\,z^3-\bar{z}^1\,z^2&z^1\,\bar{z}^2-\bar{z}^1\,z^3&z^2\,\bar{z}^2-z^1\,
 \bar{z}^1&\bar{z}^1\,\bar{z}^3-(\bar{z}^2)^2&(\bar{z}^1)^2- \bar{z}^2\,\bar{z}^3&0\cr\end{pmatrix}
\ee
Let us comment on the close resemblance of this matrix to one of the two noncommutativity matrices 
for the pure-$h$ deformation
presented in \cite{Kulaxizi:2006zc}, and in particular the one for real $h$. 
Like those matrices, the matrix $\Theta^{IJ}$ 
turns out to be independent of the radial direction $r$ when converting to spherical coordinates as in
(\ref{sphericalcoords}), and furthermore the first row and column both vanish, which guarantees that 
(a) the deformation will purely affect
the 5-dimensional internal part of the geometry, and (b) it will do so 
in a way that corresponds to a marginal deformation.  

More specifically, since (as we will also see in detail later) on taking the near-horizon limit of the full 
geometry the $r$ coordinate will become the radial direction of AdS$_5$, and will therefore (as usual in AdS/CFT) be 
associated with the scale of the gauge theory, the $r$-independence of the noncommutativity matrix 
was taken in \cite{Kulaxizi:2006zc} to be a requirement for conformal invariance of the theory. The absence
of mixing between $r$ and the remaining internal coordinates also guarantees that the geometry will neatly
split into an AdS and an internal part. The fact that our matrix satisfies these non-trivial conditions indicates
that we have indeed found the correct noncommutativity matrix beyond the holomorphic sector. 

 Let us finally note that $\Theta^{IJ}_w$ is a Poisson bivector:
\be
\Theta_w^{IL}\p_L \Theta^{JK}_w+\Theta_w^{JL}\p_L \Theta^{KI}_w+\Theta_w^{KL}\p_L \Theta^{IJ}_w=0\;.
\ee
In \cite{Kulaxizi:2006zc} this condition was seen to be necessary for associativity of the deformed star product
introduced there, as is of course also the case in the formalism of \cite{Kontsevich:1997vb}. 
Although, as discussed, we did not expect to encounter any issues with associativity for the
$w$ deformation, this provides an additional check of our formalism.

This concludes our discussion of the first-order noncommutative geometry of the $w$-deformation on the gauge theory
side. We will now turn our focus to the gravity side of the AdS/CFT correspondence, where the 
same matrix $\Theta$ will appear in the construction of the dual geometry.

\section{Generalised complex geometry} \label{GeneralisedGeometry}

In the following, we will use the noncommutative structure of the $w$-deformed gauge theory discussed in the above
to construct a gravitational
background which we will propose as its AdS/CFT dual. In order to do this most efficiently, we now need to 
introduce a second important element in our construction, namely the framework of generalised complex geometry. 

\subsection{Overview of generalised complex geometry} \label{GCGoverview}

 Generalised complex geometry, as introduced in  \cite{Hitchin03, Gualtieri04}, 
provides a framework that generalises complex and symplectic geometry while including both
as special cases. Its appeal in string theory stems from the fact that it treats the metric and B-field
on equal footing, a very natural state of affairs from the string perspective. There exist several excellent
reviews of generalised geometry aimed at physicists \cite{Grana:2005jc,Zabzine:2006uz,Koerber:2010bx}, to which 
we refer for details and references. 

For our purposes we will focus on the description of generalised geometry in terms of pure spinors, and its
relation to NS-NS supergravity backgrounds with $\Ncal=2$ supersymmetry. We follow mainly \cite{Grana:2006kf,Minasian:2006hv,Halmagyi:2007ft}. Let us start by considering three-complex-dimensional flat space $\Cset_3$, with
coordinates $z^1,z^2,z^3$ and their complex conjugates. The two relevant quantities on this manifold are the 
holomorphic volume form $\Omega$ and the K\"ahler form $J$, defined as:
\be
\Omega=\diff z^1 \wedge \diff z^2 \wedge \diff z^3 \;\quad \text{and}\quad J=\frac{i}{2}\sum_{i=1}^3 \diff z^i\wedge \diff\zb^{\bar{i}}
\ee
$\Omega$ and $J$ define a metric on the manifold, which in the holomorphic 
coordinates defined by $\Omega$ is equal to $g_{i\bar{j}}=-i J_{i\bar{j}}$. Since there is a unique six-form
on the manifold, $\Omega\wedge\overline{\Omega}$ and $J\wedge J\wedge J$ have to be related. In our 
conventions, this relation reads:
\be
J\wedge J\wedge J=\frac{3i}{4}\Omega\wedge \overline{\Omega}\;.
\ee
Since $\Cset_3$ is a complex manifold, it admits an almost complex structure, i.e. a map from the
coordinates to the coordinates which squares to minus the identity (and thus can be represented by a 
matrix with eigenvalues $\pm i$). In our case, the almost complex structure can be obtained
by raising one of the indices of $J$. This can be shown to satisfy 
$J^i_{\;k} J^k_{\;l}=-\delta^i_{\;l}$ and is thus an almost complex structure. 

Let us now move on to describe the same space in generalised complex geometry. We will consider
structures on the 12-dimensional space $T\oplus T^*$. The elements of this space (vectors $\p/\p z^I$ and
forms $\diff z^I$) will be associated to the gamma matrices $\Gamma_M$, with $M=1,\ldots,12$. Ordering the 
elements of $T\oplus T^*$ as $X=\{\iota_I,\diff z^I\}=\{\iota_1,\bar{\iota}_1,\iota_2,\ldots,\diff \zb^2,\diff z^3,\diff \zb^3\}$,
with $\iota_I$ denoting contraction, we associate
\be
\Gamma_I=\iota_I \;\quad  \text{and}\quad \Gamma_{I+6}=\diff z^I\;.
\ee
We can now define a metric on $T\oplus T^*$ using the pairing between vectors and forms:
\be
\mathcal{I}_{MN}=\twobytwo{0}{I_{6\times 6}}{I_{6\times 6}}{0}\;.
\ee
This metric, which reduces the structure group to $O(6,6)$, will be used to raise and lower indices
on the generalised tangent bundle. It is not the generalised metric, the construction of which will
be discussed momentarily.

\mparagraph{Pure spinors}

The main ingredients in the construction are certain polyforms (sums of forms of different degree) 
known as \emph{pure spinors}. The name arises because,
through the Clifford map associating forms to bispinors, they correspond to two independent pure 
spinors on the manifold (spinors annihilated by six gamma matrices). 

For flat space, the pure spinors are given by:
\be \label{purespinorsflat}
\begin{split}
\Phi^0_-&=\Omega=\diff z^1 \wedge \diff z^2 \wedge \diff z^3 \;\;\text{and}\\
\Phi^0_+&=e^{-iJ}={1\!+\!\half \sum_i\diff z^i\!\wedge\! \diff \zb^i\!
+\!\frac{1}{4}\sum_i \diff z^i\!\wedge\!\diff \zb^i\!\wedge \!\diff z^{i+1}\!\wedge\! \diff \zb^{i+1}\!+\!
\frac18 \diff z^1\!\wedge\!\diff \zb^1\!\wedge\! \diff z^2\!\wedge\! \diff \zb^2\!\wedge\! \diff z^3\!\wedge\!\diff \zb^3}  
\end{split}
\ee
which can each be seen to have an annihilator of dimension six: For $\Phi^0_-$, the annihilators are
\be
L^-_i=\diff z^i{\scriptstyle\wedge} \;,\quad \text{and}\quad L^-_{\bar{i}}=\iota_{\bar{i}}
\ee
while for $\Phi^0_+$ they are
\be
L^+_i=\diff z^i{\scriptstyle\wedge}+2\iota_{\bar{i}}\;,\quad \text{and}\quad L^+_{\bar{i}}=\diff \bar{z}^{\bar{i}}{\scriptstyle\wedge}-2\iota_i\;.
\ee
For these pure spinors to define an $\Ncal=2$ background, they also need to be closed and 
satisfy the compatibility conditions \cite{Halmagyi:2007ft}:
\be \label{compatibilityone}
\langle \Phi_-,X\Phi_+\rangle=0 \;,\quad \langle\bar{\Phi}_-,X\Phi_+\rangle=0
\ee
with $X$ being any element of the tangent or cotangent space (as above), and 
\be \label{compatibilitytwo}
\langle\bar{\Phi}_+,\Phi_+\rangle=\langle\bar{\Phi}_-,\Phi_-\rangle
\ee
where we made use of the Mukai pairing on the space of generalised forms:
\be \label{innerproduct}
\langle A, B\rangle=\sum (-1)^{[n/2]} A_n\wedge B_{6-n}\;.
\ee
Here $A_n,B_{6-n}$ are the components of $A$ and $B$ of the corresponding degree 
and $[\cdot]$ denotes the integer part.

The compatibility  conditions can be easily verified for the flat-space pure spinors in (\ref{purespinorsflat}). This
is to be expected, as flat space is a (very special) case of a $\Ncal=2$ background.

\mparagraph{From pure spinors to NS-NS fields}

Of course, a compatible pair of pure spinors is a rather implicit encoding of the metric and B-field of the 
solution. To obtain a more explicit construction of  
the metric and B-field one needs to consider the generalised complex structures associated to the
pair of pure spinors. 

Using the mapping of the gamma matrices to elements of $T\oplus T^*$, we can define the antisymmetric combinations $\Gamma_{MN}$  as
\be
\begin{split}
\Gamma_{I,J}&=\half(\iota_I\iota_J-\iota_J\iota_I)=\iota_I\wedge \iota_J\;,\;\quad 
\Gamma_{I,J+6}=\half(\iota_I \diff z^J-\iota_J \diff z^I)\;,\\
\Gamma_{I+6,J}&=\half(\diff z^I \iota_J-\diff z^J \iota_I)\;,\;\quad
\Gamma_{I+6,J+6}=\half(\diff z^I\diff z^J-\diff z^J\diff z^I)=\diff z^I\wedge \diff z^J\;. 
\end{split}
\ee
The generalised complex structures can now be obtained through the Mukai pairing as 
\be \label{generalisedcomplex}
\Jcal_{\pm\;MN}=\langle \overline{\Phi}_{\pm},\Gamma_{MN}\Phi_{\pm}\rangle\;.
\ee
The generalised complex structures square to $-1$, that is $\Jcal_{\pm}{}^M_{\;N}\Jcal_{\pm}{}^N_{\;R}=-\delta^{M}_{\;R}$.\footnote{Depending on conventions, a normalisation constant might need to be included in (\ref{generalisedcomplex}) 
for this to happen.}
Equivalently, one can construct the 
generalised complex structures through the annihilator space of the pure spinors. In particular, they should
be such that the corresponding annihilator vectors $L^\pm$ (see above) form their $\pm i$ eigenspaces respectively.

Compatibility of the pure spinors guarantees that the generalised complex structures commute,
\be
\Jcal_+{}^{M}_{\;N}\Jcal_-{}^{N}_{\;L}-\Jcal_-{}^{M}_{\;N}\Jcal_+{}^{N}_{\;L}=0
\ee
which allows us to unambiguously define the (12-dimensional) generalised metric as:
\be \label{generalisedmetricdef}
\Gcal_{MN}=- \Jcal_{+ML}\Jcal^{\;L}_{-\;\;N}\;.
\ee
The generalised metric squares to one and can be shown \cite{Gualtieri04} to be related to the six-dimensional metric and B-field as\footnote{It is required to raise the first index in
order to bring the metric into this canonical form. Leaving it down would just permute the $6\times 6$ blocks.}
\be \label{generalisedmetric}
\Gcal^M_{\;N}=\twobytwo{-g^{-1}B}{g^{-1}}{g-Bg^{-1}B}{Bg^{-1}}
\ee
while in our conventions the dilaton will be given by
\be \label{dilaton}
e^{2\Phi}=\sqrt{|\det g|} \;
\ee
in accordance with $O(6,6)$ invariance (see \cite{Grana:2008yw} for relevant comments).\footnote{In 
\cite{Halmagyi:2007ft}, the dilaton satisfies $e^{-2\Phi}=({\Phi}_\pm,\overline{\Phi}_\pm)$, with the norm
defined as $({\Phi}_\pm,\overline{\Phi}_\pm)=\langle {\Phi}_\pm,\overline{\Phi}_\pm\rangle/\text{vol}_g$, with
$\text{vol}_g$ the volume form of the metric $g$. All our pure spinors will satisfy 
$\langle {\Phi},\overline{\Phi}\rangle=1$, hence the equality in (\ref{dilaton}). } 
We have thus obtained the three NS-NS fields of our background starting from the pure spinors. 

Going through the above steps for the flat-space case (\ref{purespinorsflat}), we find 
\be
\Gcal=\twobytwo{0}{g^{-1}}{g}{0}\;,\quad\text{where}\quad 
g= \begin{pmatrix}
0&1&0&0&0&0\\
1&0&0&0&0&0\\
0&0&0&1&0&0\\
0&0&1&0&0&0\\
0&0&0&0&0&1\\
0&0&0&0&1&0
\end{pmatrix}
\ee
which corresponds to the flat-space complex metric $\diff s^2=g_{i\bar{j}} \diff z^i\diff \zb^{\bar j}$ and
vanishing B-field and dilaton.

\subsection{Twisting the pure spinors}

We will now consider deforming the flat 6-dimensional geometry above in order to find new NS-NS solutions. 
The deformations we will consider are bivector-type deformations, which are of the form
\be \label{bivectordeformation}
\Phi_{\pm}=e^{\beta^{IJ}\iota_I\wedge \iota_J}\Phi_{\pm}^0
\ee 
Such deformations are known to be equivalent to TsT transformations in cases with isometries \cite{Minasian:2006hv,
Halmagyi:2007ft,Andriot:2009fp}.\footnote{$\beta$-transformations and their relation with T-duality 
also feature crucially in the formalism of $\beta$-supergravity \cite{Andriot:2011uh,Andriot:2013xca,Andriot:2014qla}
which describes backgrounds with non-geometric fluxes.}
 We would like to eventually apply them also to more general cases to produce compatible pure spinor pairs 
corresponding to a given deformed gauge theory. Our claim will be that for the integrable cases 
the bivector $\beta^{IJ}$ above can simply be taken to be the noncommutativity matrix $\Theta^{IJ}$. 

In line with the noncommutative flavour of this work, we will also advocate an alternative way 
of thinking about the above deformation, which is equivalent to first order in the deformation parameter
but might provide an alternative description at higher orders. We will propose that it can be written 
as a \emph{non-anticommutative} deformation, by introducing a star-product on the space of polyforms: 
\be
\diff z^I \wedge_\ast \diff z^J=\left(1-\frac{i}2\Theta^{KL}\iota_K\wedge\iota_L\right) \diff z^I\wedge \diff z^J=
\diff z^I \wedge\diff z^J-i\Theta^{IJ}
\ee
where $\Theta^{IJ}$ now plays the role of a non-anticommutativity matrix.\footnote{The minus in the 
definition is just a choice of sign for the deformation parameter,
 but the $i$ is important for compatibility of the deformed pure spinors.
See comments in \cite{vanTongeren:2015uha} for the need to introduce an $i$ in a related context.}
 So the star product deforms
the anticommutative wedge product to a generically non-anticommutative one. This is in a similar
spirit as in \cite{Kulaxizi:2004pa}, where a very similar non-anticommutative star product was 
introduced between the fermionic coordinates of supertwistor space \cite{Witten:2003nn} to compute
amplitudes in the general Leigh-Strassler gauge theory. 

The relation between bivector deformations and noncommutativity is well established, especially since the work
of Kontsevich \cite{Kontsevich:1997vb}. In the generalised geometry context, it is also well known that 
bivector deformations in the presence of isometries lead to new backgrounds which can also be obtained 
by a succession of T 
dualities\footnote{This is unlike B-twists, which are symmetries of the generalised metric and do not 
lead to new solutions.}. Relevant discussions can be found in \cite{Ellwood:2006ya,Jurco:2013upa}. 
So in the case where there are isometries to T-dualise along, it is not a surprise that a TsT deformation 
is equivalent to a bivector deformation. Our proposal is that the noncommutative structure of the gauge 
theory can guide us as to which bivector twist generates its dual geometry also  in the absence 
of isometries.  

 Previous work on star products on forms can be found in \cite{McCurdy:2008ew,McCurdy:2009xz} as
well as the recent work \cite{Arias:2015wha} which deals specifically with differential Poisson algebras. 
The main novelty in our case is that, since we are working in the framework of generalised geometry, our star
products will act on polyforms and will mix the degrees of the forms involved. In particular, each action
of the star product will result in a component of degree lower by two in the polyform. 
 
 Concretely, we propose that, for a given field-theoretical deformation parametrised by a noncommutativity
matrix $\Theta^{IJ}$, the corresponding deformation on the supergravity side is obtained by considering the flat-space
pure spinors, but multiplied with the above star product instead of the wedge product. So we obtain:\footnote{This
is a minimal non-anticommutative extension of the usual wedge product, in the sense that all terms arise
by introducing a star product between existing terms. One could also consider additional
terms which tend to zero in the undeformed limit, but we have so far not seen the need for such terms.}
\be \label{deformedpurespinors}
\begin{split}
\Phi_-^{\ast}&=\diff z^1\wedge_\ast \diff z^2 \wedge_\ast \diff z^3 \;\; \text{and}\\
\Phi_+^{\ast}&=1\!+\!\half \sum_{i=1}^3\diff z^i\wedge_\ast \diff \zb^i
\!+\!\frac{1}{4}\sum_{i=1}^3\diff z^i\wedge_\ast\diff \zb^i\wedge_\ast \diff z^{i+1}\wedge_\ast \diff \zb^{i+1}\\
&\quad\; +\!\frac18 \diff z^1\wedge_\ast\diff \zb^1\wedge_\ast \diff z^2\wedge_\ast \diff \zb^2\wedge_\ast 
\diff z^3\wedge_\ast\diff \zb^3
\end{split}
\ee
In the following we will first apply this procedure to the $\beta$-deformation, in order to recover the 
NS-NS precursor of the LM metric \cite{Lunin:2005jy,Halmagyi:2007ft}. After this check, we will
proceed to the $w$-deformation and find the same metric, but in a form more adapted to the $w$-deformed gauge theory.

 An important comment is in order, however. From the above it might appear that we are introducing some kind of 
non-anticommutativity on the gravity side, which
might be taken to imply that the closed-string metric (in the framework of \cite{Seiberg:1999vs}) is somehow becoming
non-commutative. We hasten to emphasise, however, that for the deformations we will consider the 
procedure (\ref{deformedpurespinors})
is completely equivalent to the well-known bivector deformation (\ref{bivectordeformation}). The reason is simply
that (as can be checked) only the first-order part in the expansion (\ref{bivectordeformation}) is nonzero, and that 
for our deformations the star product reduces to doing all single Wick contractions in (\ref{deformedpurespinors}),
with double- and higher Wick contractions vanishing.  
This property is very specific to our deformations and is probably linked to quasi-triangularity. Future studies
of non-quasitriangular deformations will decide whether the star-product formulation is useful beyond leading order.

\subsection{The $\beta$-deformation in generalised complex geometry}

As shown in \cite{Halmagyi:2007ft}, the NS-NS precursor of the $\beta$-deformation can be obtained through a 
bivector transformation
on the pure spinors. It can easily be checked that the transformation performed in \cite{Halmagyi:2007ft} is
equivalent to introducing the following star-product relations
\be
\diff z^i \wedge_\ast \diff z^j=\diff z^i \wedge \diff z^j-i\Theta^{ij}\;,\;\quad 
\diff z^i \wedge_\ast \diff \zb^{\bar{j}}=\diff z^i \wedge\diff \zb^{\bar{j}}-i\Theta^{i\bar{j}} 
\ee
and their conjugates, with the noncommutativity matrix being (\ref{betaTheta}).

In \cite{Kulaxizi:2006pp}, this matrix was combined with the Seiberg-Witten technique of constructing
a closed-string metric with a B-field from a non-commutative open string metric to reproduce the NS-NS
fields of the LM metric. (For earlier work in the same direction, see \cite{CatalOzer:2005mr}). 

Let us use this matrix to deform the 6-dimensional
flat-space pure spinors as discussed in the previous section. It turns out that we only need to consider
single Wick contractions (leading to linear terms in $\beta$). A straightforward computation results
in the following  beta-deformed pure spinors:
\be \label{betadeformedpurespinors}
\begin{split}
\Phi_-^\beta&=\diff z^1\wedge \diff z^2\wedge \diff z^3+\beta ~\diff (z^1z^2z^3)\quad \text{and}\\
\Phi_+^\beta&=\Phi_+^0-\frac{\beta}{4}\left[\zb^1\zb^2\diff z^1\wedge \diff z^2 +\zb^1 z^2 \diff z^1\wedge \diff \zb^2
+z^1 \zb^2\diff \zb^1\wedge\diff z^2+z^1z^2\diff \zb^1\wedge\diff \zb^2 \;\;\text{+cyclic}\right]
\end{split}
\ee
This is the same result as that in \cite{Halmagyi:2007ft}, which, as mentioned, 
applied an equivalent bivector deformation to the flat-space pure spinors\footnote{ 
In \cite{Halmagyi:2007ft}, a B-transformation was employed to keep $\Phi_+$ equal to its undeformed value. That 
transformation does not affect the final answer for the supergravity fields, so for simplicity we choose not to 
perform it here.} 
Applying the procedure outlined in (\ref{GCGoverview}) to obtain the supergravity NS-NS fields we can write the
metric as\footnote{We have divided the metric by two to revert to the canonical normalisation 
where $g_{i\bar{i}}=\half$ for flat space.}
\be
\diff s^2= \frac{G}{4}\left(g_{i,i} \diff z^i\diff z^i+g_{i,i+1} \diff z^i\diff z^{i+1}+g_{i,\bar{i}} \diff z^i\diff \zb^{\bar{i}}+g_{i,\overline{i+1}} \diff z^i\diff \zb^{\overline{i+1}} +\text{c.c.}\right)
\ee
where the components are
\be \label{LMgcomplex}
\begin{split}
g_{i,i}&=\beta^2 (\zb^{\bar{i}})^2(z^{i+1}\zb^{\overline{i+1}}+z^{i-1}\zb^{\overline{i-1}})\;,\\
g_{i,\bar{i}}&=(2+\beta^2(z^i\zb^{\bar{i}}(z^{i+1}\zb^{\overline{i+1}}+z^{i-1}\zb^{\overline{i-1}})+2z^{i-1}\zb^{\overline{i-1}}z^{i+1}\zb^{\overline{i+1}})\;,\\
g_{i,i+1}&=\beta^2 \zb^{\bar{i}}\zb^{\overline{i+1}} z^{i-1}\zb^{\overline{i-1}}\;, \;\;
g_{i,\bar{i+1}}=\beta^2 \zb^{\bar{i}} z^{i+1} z^{i-1}\zb^{\overline{i-1}}\;,
\end{split}
\ee
with $G^{-1}=1+z^1\zb^1z^2\zb^2+z^2\zb^2z^3\zb^3+z^3\zb^3z^1\zb^1$, while the remaining components are determined by symmetry. 
The independent B-field components are:
\be
B_{i,i+1}=G \beta \zb^{\bar{i}}\zb^{\overline{i+1}}/4 \quad\text{and} \quad B_{i,\overline{i+1}}=-G\beta \zb^{\bar{i}}z^{i+1}/4
\ee
while the dilaton is $e^{2\Phi}=G$. Converting to real coordinates we recover  the $\Ncal=2$ background 
discussed in the appendix of \cite{Lunin:2005jy}, the NS-NS precursor
of the LM background. Adding D3-branes at the origin of this geometry and taking the near-horizon limit 
leads to the $\Ncal=1$ real-$\beta$ LM solution. As these solutions are well known, we do not reproduce them here. 
In the following we will apply the same techniques to the $w$-deformation and will provide more details of the
steps to the $\Ncal=1$ solution to highlight the features one might expect to encounter in the generic case.

The $\Ncal=1$ LM solution has also been obtained in the generalised geometry context directly (without
going through the NS-NS precursor) by considering the action of T-duality on the pure 
spinors of AdS$_5\times \Srm^5$ \cite{Minasian:2006hv}. In the following we will follow the two-step 
process above for the $w$-deformation, the main appeal being that we start with the pure spinors of flat space, 
which is closest to the quantum plane picture we developed on the gauge theory side. It would definitely be
worth exploring whether the kinds of deformations that we are interested in can be formulated as an action on 
the pure spinors of AdS$_5\times \Srm^5$.

\subsection{The $w$-deformation in generalised complex geometry} \label{wgcgeometry}

We are now ready to consider the action of the $w$-deformed star product on the flat-space pure spinors. Using
(\ref{deformedpurespinors}) with the noncommutativity matrix (\ref{wThetamatrix}) we obtain
\be
\Phi_-^w=\diff z^1\wedge \diff z^2\wedge \diff z^3-i w [(z^2z^3-(z^1)^2)\diff z^1+(z^3z^1-(z^2)^2)\diff z^2+(z^1z^2-(z^3)^2)\diff z^3]
\ee
and 
\be
\begin{split}
\Phi_+^w&=\Phi_+^0
+\frac{i w}{4} \left[(z^3\zb^3-z^2\zb^2)\diff z^1\wedge \diff \zb^1 
+ ((\zb^3)^2-\zb^1\zb^2) \diff z^1\wedge\diff z^2
+ (z^1\zb^3-\zb^2 z^3) \diff z^1\wedge\diff \zb^2 \right.\\
&\left.\quad\qquad\qquad\;+(z^2\zb^3-z^1\zb^2)\diff z^1\wedge\diff \zb^3+(z^1z^2-(z^3)^2)\diff\zb^1\wedge \diff \zb^2 \;\text{+cyclic}\right]\\
& \quad\quad\;\;+\frac{iw}{8}\left[(z^1\zb^1-z^2\zb^2)\diff z^1\wedge\diff \zb^1\wedge\diff z^2\wedge\diff \zb^2
-(\zb^2\zb^3-(\zb^1)^2)\diff z^1\wedge \diff \zb^1\wedge \diff z^2\wedge \diff z^3 \right.\\
&\qquad\qquad\quad\; +(\zb^1 z^3-z^1\zb^2)\diff z^1\wedge\diff \zb^1\wedge\diff \zb^2\wedge\diff \zb^3
-(z^1\zb^3-\zb^1 z^2)\diff z^1\wedge \diff \zb^1\wedge \diff z^2\wedge \diff \zb^3\\
&\left. \qquad\qquad\quad\; +(z^2z^3-(z^1)^2)\diff z^1\wedge \diff \zb^1\wedge \diff \zb^2\wedge \diff \zb^3\;\text{+cyclic}\right]
\end{split}
\ee
We will now show that these forms are pure spinors, and are also compatible. They 
thus define an $\Ncal=2$ NS-NS background. 
To do this, we first construct the annihilators of $\Phi_\pm^w$:
\be
\begin{split}
L^-_{(1)}&=\diff z^1 -i w \left[(z^1z^2-(z^3)^2)\iota_2-(z^1z^3-(z^2)^2) \iota_3\right]\;,\\ 
L^-_{(\bar{1})}&=\iota_{\bar{1}}\;
\end{split}
\ee
and 
\be
\begin{split}
L^+_{(1)}&=\diff z^1  +\left[\left(2-iw(z^3\zb^3-z^2\zb^2)\right)\iota_{\bar{1}}-iw(z^3\zb^2-z^2\zb^1)\iota_{\bar{3}}\right.\\
&\qquad\qquad \left.+iw(z^2\zb^3-z^3\zb^1)\iota_{\bar{2}}+iw(z^1z^3-(z^2)^2)\iota_3-iw(z^1z^2-(z^3)^2)\iota_2\right]\;,\\
L^+_{(\bar{1})}&=\diff \zb^1+ \left[\left(-2+iw(z^3\zb^3-z^2\zb^2)\right)\iota_1+iw(z^2\zb^3-z^1\zb^2)\iota_3-iw(z^3\zb^2-z^1\zb^3)\iota_2\right. \\
&\qquad\qquad \left.-iw(\zb^1\zb^3-(\zb^2)^2)\iota_{\bar{3}}+iw(\zb^1\zb^2-(\zb^3)^2)\iota_{\bar{2}}\right]\;.
\end{split}
\ee
plus their cyclic permutations. The existence of six annihilators each confirms that $\Phi_+$ and $\Phi_-$ are
pure spinors. It is also easy to see that they share three annihilators. Concretely, since we can write
the annihilator $L^+_{(1)}$ as
\be
L^+_{(1)}=L^-_{(1)}-iw(z^3\zb^3-z^2\zb^2)L^-_{(\bar{1})}-iw(z^3\zb^2-z^2\zb^1) L^-_{(\bar{3})}+iw(z^2\zb^3-z^3\zb^1)L^-_{(\bar{2})}
\ee
it also annihilates $\Phi^w_+$, as do its two cyclic permutations. This establishes the compatibility of the two pure spinors.\footnote{Alternatively, one can also check the compatibility conditions (\ref{compatibilityone}) and 
(\ref{compatibilitytwo}), which are also satisfied.}

 We can also straightforwardly check that $\Phi_\pm$ are closed and that they have equal Mukai norms:
\be
\langle {\Phi}_-,\overline{\Phi}_-\rangle=\langle {\Phi}_+,\overline{\Phi}_+\rangle=1
\ee
which are equal to those of the undeformed $\Cset_3$. 
Together, these conditions tell us that we have an $\Ncal=2$ background \cite{Halmagyi:2007ft}, the NS-NS 
precursor of the dual to the $w$-deformation.

\section{The $w$-twisted background} \label{Metric}

Although the generalised complex description above is in principle sufficient, for many applications (especially
 to AdS/CFT)
it is desirable to express the background in terms of the supergravity fields. To do this, we will first 
construct the generalised metric, from which it is easy to read off the supergravity fields. 

Afterwards, we will write down the $w$-deformed solution in real coordinates. We will start with the 10-dimensional
NS-NS precursor solution, which is simply the direct product of 4d Minkowski space with the six-dimensional deformation
of flat space defined above. We will then consider the R-R solution which would result from the addition of D3-branes
to this background and taking the near-horizon limit. For the $w$-deformation it takes 
the form of AdS$_5\times \Srm_w^5$, where $\Srm_w^5$ is a deformation of the 5-sphere. A coordinate redefinition
shows that this sphere is nothing but the LM metric in an unusual coordinate system. This is in line with our
field theory expectations, providing a check of the validity of our approach and the appropriate application
of the twist.

\subsection{The metric in complex coordinates}

Our first task is to construct the generalised complex structures following (\ref{generalisedcomplex}), taking
care to normalise them appropriately in order for them to square to $-1$. Their expressions can be found 
in appendix \ref{GCS}. We now 
obtain the generalised metric according to (\ref{generalisedmetricdef}). Using (\ref{generalisedmetric}),
we can easily read off the string-frame metric, B-field and dilaton of our NS-NS background. The end 
result of this procedure  can be exhibited more clearly by defining the factor:
\be
\begin{split}
G^{-1}=1+w^2&\left[z_1^2\zb_1^2+z_2^2\zb_2^2+z_3^2\zb_3^2+z_1\zb_1z_2\zb_2+z_2\zb_2z_3\zb_3+z_1\zb_1z_3\zb_3\right.\\
&\left.-z_1z_2\zb_3^2-z_2z_3\zb_1^2 -z_3z_1\zb_2^2-z_1^2\zb_2\zb_3-z_2^2\zb_3\zb_1-z_3^2\zb_1\zb_2\right]
\end{split}
\ee
which is the inverse of the square root of the determinant of the metric.\footnote{The determinant of the 
metric in complex coordinates is $-G^{-2}$, but will become positive on transformation to real coordinates.}
We then find that the six-dimensional part of the (string-frame) metric is given by
\be
\diff s^2=\frac{G}4~\left[g_{i,i}\diff z^i\diff z^i+g_{i,\bar{i}}\diff z^i\diff \zb^{\bar{i}}+g_{i,i+1}\diff z^i\diff z^{i+1}
+g_{i,\overline{i+1}}\diff z^i\diff \zb^{\overline{i+1}}+\text{c.c.}\right]
\ee
with the components given by\footnote{We have placed the indices down and removed the bars from conjugate indices for increased readability.}
\be  \label{wdefgcomplex}
\begin{split}
g_{i,i}=&w^2[ z_i(\zb_{{i+1}}^3+\zb_{{i-1}}^3)+\zb_{{i}}(z_{i+1}\zb_{{i-1}}^2+z_{i-1}\zb_{{i+1}}^2)
-\zb_{{i+1}}\zb_{{i-1}}(z_{i-1}\zb_{{i-1}}+z_{i+1}\zb_{{i+1}})-2~z_i\zb_i\zb_{{i+1}}\zb_{i-1} ]\;,\\
g_{i,\overline{i+1}}=&w^2(z_i^2-z_{i+1}z_{i-1})(\zb_{{i+1}}^2-\zb_{{i+2}}\zb_{i})\;,\\
g_{i,i+1}&=w^2[\zb_i^2(z_i\zb_{i+1}-z_{i+1}\zb_{i-1})+\zb_{i+1}^2(z_{i+1}\zb_{i}-z_i\zb_{i-1})+z_{i-1}\zb_{i-1}(\zb_{i-1}^2-\zb_i\zb_{i+1})]\\
g_{i,\bar{i}}=&2+w^2\left(2z_i^2\zb_i^2+z_{i+1}^2\zb_{i+1}^2+z_{i-1}^2\zb_{i-1}^2
+2z_{i+1}\zb_{i+1}z_{i-1}\zb_{i-1} + z_i\zb_i(z_{i-1}\zb_{i-1}+z_{i+1}\zb_{i+1})\right.\\
&\left.-(2z_i^2\zb_{i+1}\zb_{i-1}+z_{i+1}^2\zb_{i}\zb_{i-1}+z_{i-1}^2\zb_i\zb_{i+1}+\text{c.c.})\right)\;,\\
g_{\bar{i}\bar{j}}=&\overline{g_{ij}}\;,\quad g_{\bar{j},i}=\overline{g_{i\bar j}}\;.
\end{split}
\ee
while the B-field is
\be
B=\half \left(B_{i\bar{i}}\diff z^i\wedge\diff \zb^{\bar{i}}+B_{i,i+1}\diff z^i\wedge \diff z^{i+1}+B_{i,i-1}\diff z^i\wedge \diff z^{i-1}+B_{i,\overline{i+1}}\diff z^i\wedge \diff \zb^{\overline{i+1}}+\text{c.c.}\right)
\ee
with
\be
\begin{split}
&B_{i\bar{i}}=-B_{\bar{i}i}=\frac{i w}{4} G \left(z_{i-1}\zb_{i-1}-z_{i+1}\zb_{i-1}\right)\\
&B_{i,i+1}=-B_{i+1,i}=\frac{i w}{4} G \left(\zb_i\zb_{i+1}-\zb_{i-1}^2\right)\\
&B_{i,\overline{i+1}}=-B_{\overline{i+1},i}=\frac{i w}{4}\left(z_i\zb_{i-1}-z_{i-1}\zb_{i+1}\right)
\end{split}
\ee
Finally, the dilaton is simply 
\be
e^{2\Phi}=G\;.
\ee
Already at this level it can be seen that the metric (\ref{wdefgcomplex}) is nothing but the LM precursor 
metric \cite{Lunin:2005jy} written in complex
coordinates. Transforming the $z^i,\zb^i$ coordinates of (\ref{LMgcomplex}) as 
\be \label{zredefinition}
z^i\ra (T^\dag)^i_{\;j}z^j \;\quad,\quad  \zb_i\ra \zb_j T^j_{\;i}
\ee
with $T$ in (\ref{Tmatrix}), and with an additional rescaling $\beta^2=3w^2$ (which arises from the rescaling needed
to precisely match the $r$-matrices in the two cases, c.f. section \ref{wdeffirstorder}), we easily reproduce (\ref{wdefgcomplex}). 

In the following section we will rewrite these fields in real coordinates to facilitate comparison 
with the more usual parametrisation of the $\beta$-deformation \cite{Lunin:2005jy}, as well as 
the perturbative solution of \cite{Kulaxizi:2006zc} for the $h$-deformation.

\subsection{The NS-NS precursor solution} \label{Precursor}

Let us start by converting the $w$-deformed NS-NS metric to real coordinates. We will use the coordinates:
\be
z^k=r_k e^{i\phi_k}\;,\quad \zb^k=r_k e^{-i\phi_k}\;.
\ee
As discussed, we expect the $w$-deformed metric to only have one obvious $\Urm(1)$ isometry, the one corresponding
to the $\Urm(1)$ R-symmetry of the corresponding gauge theory. In the above coordinates this symmetry 
is $\phi_i\ra \phi_i+a$. We thus expect the angle dependence of the metric components to be in terms
of combinations invariant under this transformation. It turns out that the needed combinations arise in
four distinct forms, for which we define the following shorthand notation:\footnote{A similar notation was used 
in \cite{Kulaxizi:2006zc}.}
\be
\begin{split}
C_i&=\cos(2\phi_i-\phi_{i+1}-\phi_{i-1})\;,\quad C_{i,j}=\cos(3(\phi_i-\phi_{j}))\\
S_i&=\sin(2\phi_i-\phi_{i+1}-\phi_{i-1})\;,\quad S_{i,j}=\sin(3(\phi_i-\phi_{j}))\;.\\
\end{split}
\ee
The $G$ parameter now takes the form:
\be
G^{-1}=1\!+\!w^2(r_1^4+r_2^4+r_3^4+r_2^2 r_3^2+r_1^2r_3^2+r_1^2r_2^2-2C_3~r_1r_2r_3^2-2C_2~r_3r_1r_2^2-2C_1~r_2r_3r_1^2)
\ee
Going to Einstein frame, the full 10-dimensional metric is 
\be
\diff s^2=G^{-\frac14}\left(-\diff t^2+\sum_{i=1}^3\diff x_i^2+G\left( g^{rr}_{ij} \diff r_i\diff r_j
+g^{r\phi}_{ij}\diff r_i\diff\phi_j+g^{\phi\phi}_{ij}\diff \phi_i\diff\phi_j\right)\right)
\ee
with the $rr$ components being
\be
\begin{split}
g^{rr}_{11}&=1+\frac{w^2}{2}\left(2r_1^4+r_2^4+r_3^4+r_1^2r_2^2+r_1^2r_3^2+r_2^2r_3^2-C_{31}r_1r_3^3-C_{12}r_1r_2^3\right.\\ 
&\;\quad\quad\left.+C_1(r_2r_3^3+r_2^3r_3-2 r_1^2r_2r_3)-3C_2r_1r_2^2r_3-3C_3r_1r_2r_3^2\right)\\
g^{rr}_{12}&=\frac{w^2}2\left(r_1 r_2^3+r_2r_1^3+C_{1,2}r_1^2r_2^2+C_3~r_3^4-C_2~(r_2^3r_3+r_1^2r_2r_3)
-C_1(r_1^3r_3+r_1r_2^2r_3)\right)\;,
\end{split}
\ee
the $r\phi$ components (note that $g^{\phi r}=g^{r\phi}$)
\be
\begin{split}
g^{r\phi}_{11}&=-\frac{w^2}2 r_1 \left( S_1 (r_2 r_3^3+r_2^3r_3+2 r_1^2r_2r_3)+S_2 r_1r_2^2r_3+S_3r_1r_2r_3^2+S_{3,1}r_1r_3^3-S_{1,2}r_1 r_2^3\right)\\
g^{r\phi}_{12}&=-\frac{w^2}2 r_2 \left(S_1(r_1^3r_3-r_1r_2^2r_3) -S_3 r_3^4-S_2(r_2^3r_3+r_1^2r_2r_3)-S_{1,2}r_1^2r_2^2\right)\\
g^{r\phi}_{13}&=-\frac{ w^2}2 r_3 \left(S_1(r_1^3r_2-r_1r_2r_3^2) -S_2 r_2^4-S_3(r_2r_3^2+r_1^2r_2r_3)+S_{3,1}r_1^2r_3^2\right)
\end{split}
\ee
and the $\phi\phi$ components:
\be
\begin{split}
g^{\phi\phi}_{11}&=r_1^2+\frac{w^2}2r_1^2\left(2r_1^4+r_2^4+r_3^4+r_1^2r_2^2+r_1^2r_3^2+2r_2^2r_3^2\right.\\
&\;\quad\quad\left.-C_1(r_2r_3^3+r_2^3r_3-6r_1^2r_2r_3)-C_2r_1r_2^2r_3-C_3r_1r_2r_3^2+C_{1,2}r_1r_2^3+C_{3,1}r_1r_3^3\right)\\
g^{\phi\phi}_{12}&=\frac{w^2}2 r_1r_2\left(2r_1r_2r_3^2-r_1r_2^3-r_1^3r_2-C_3r_3^4+C_2(r_1^2r_2r_3-r_2^2r_3)
+C_1(r_1r_2^2r_3-r_1^3r_3)+C_{1,2}r_1^2r_2^2\right)\\
\end{split}
\ee
with the remaining components fixed by cyclicity ($g^{rr}_{12}=g^{rr}_{23}=g^{rr}_{31}$ etc.) and symmetry of the metric. The B-field components take the form:
\be
\begin{split}
B_{r_1,r_2}&=-G w r_3(r_1 S_1 +r_2 S_2+r_3 S_3)/2\\
B_{r_1,\phi_1}&=G w r_1(r_3^2-r_2^2)/2\\
B_{r_1,\phi_2}&=G w r_2(-r_1r_2 +C_3 r_3^3-C_2 r_2r_3+C_1r_1r_3)/2\\
B_{r_1,\phi_3}&=G w r_3(r_1r_3+C_3r_2r_3-C_2 r_2^2-C_1 r_1r_2)/2\\
B_{\phi_1,\phi_2}&=Gw r_1r_2r_3(r_3 S_3-r_2 S_2-r_1 S_1)/2\\
\end{split}
\ee
(with the remaining components determined by cyclicity and antisymmetry) while the dilaton is
\be
e^{2\Phi}=G
\ee
as usual. 

It can be checked explicitly that the above fields provide a solution of the IIB supergravity equations (which 
we record in appendix \ref{IIBEOM} for completeness). This is the NS-NS precursor of the $w$-deformed 
background. In the next section we will switch to spherical coordinates, in which the solution can be seen
to take the form of a cone over a five-dimensional geometry.

\subsection{The R-R solution} \label{RRsolution}

  The final step in our construction of the dual $w$-deformed geometry is to add D3-branes to the NS-NS precursor 
solution
and take the near-horizon limit, which will lead to a R-R solution with an AdS and internal deformed sphere part. 
This is straightforward to do, since if we parametrise the $r_i$ coordinates as (c.f. (\ref{sphericalcoords}))
\be
r_1=r\cos\alpha\;,\quad r_2=r\sin\alpha\sin\theta\;,\quad r_3=r\sin\alpha\cos\theta
\ee
the six-dimensional part of the metric takes the form 
\be \label{cone}
\diff s^2=\diff r^2+r^2\diff \Omega^2_w\;.
\ee
We see that the precursor takes the form of a cone with radial parameter $r^2=r_1^2+r_2^2+r_3^2$ 
over a 5-dimensional base, with $\diff \Omega^2_w$ denoting the metric of a deformed sphere forming 
the base of the cone. The RR background we are aiming to construct arises  by placing
a stack of $N$ D3-branes at $r=0$ and taking the near-horizon limit. 
Exactly as in the case of $\Srm^5$ or the LM metric, the near-horizon limit will result
in $r$ becoming the AdS$_5$ radial direction, leaving us with a 5-dimensional compact metric $\Srm^5_w$
parametrised by the angles $\alpha,\theta,\phi_1,\phi_2,\phi_3$. 
This AdS$_5\times\Srm^5_w$ is expected to be an $\Ncal=1$ solution of IIB supergravity and it is natural
to identify it with the AdS/CFT dual of the $w$-deformation. 
 
 We will not go through the process of finding the D3-brane solution and taking the near-horizon limit,
but will take the shortcut of constructing the R-R fields of the near horizon background by solving 
the supergravity equations of motion, applying a fair amount of intuition from the known cases. In 
this we follow \cite{Kulaxizi:2006zc}, which used the same
approach to construct the dual to the $h$ deformation to third order. The limitation there was caused 
by ambiguities related to non-associativity of the star product used to derive the NS-NS solution. As discussed, 
our ability to control associativity led us to the exact NS-NS background for the $w$-deformation. The 
construction of the all-orders RR background in our case will therefore be straightforward, although its verification
will be computationally demanding because of the complexity of the fields involved. 

Of course, the generalised geometry framework encompasses AdS$_5$ $\Ncal=1$ 
backgrounds as well (see e.g. \cite{Grana:2005sn,Gabella:2009wu,Gabella:2010cy}), and it would be possible 
(and would certainly provide further insight) to take the near-horizon 
limit directly at the pure spinor level and express the R-R background in generalised geometry language. 
We leave this construction to future work. 

 In this section it will be convenient to switch back to the string frame, where the 10-dimensional metric 
will simply be a direct product AdS$_5\times\Srm^5_w$.  The AdS$_5$ part is standard, so we will just 
focus on the geometry of the deformed five-sphere. From (\ref{cone}) it is clear that the metric of the
sphere will just be the five-dimensional part of the 6d metric written in spherical coordinates, with $r$
replaced by $R$, the radius of AdS$_5$. The 5d $B$-field will also be derived in the same way. 
The remaining ingredient to obtain the RR solution is the introduction
of the RR form fields. The five-form flux will take the standard self-dual form, and the RR zero-form will be taken
to vanish. So the only nontrivial flux will be that of the RR two-form, which will be obtained from the 
B-field using the equations of motion. 
In the following we will just list the outcome of this procedure. 

In our spherical coordinates, the $G$-factor takes the form:
\be
G^{-1}=1\!+\!w^2R^4(1-s_\alpha^2c_\alpha^2-s_\alpha^4s_\theta^2c_\theta^2-2c_\alpha s_\alpha^3 c_\theta s^2_\theta C_2
-2s_\alpha^2c_\alpha^2s_\theta c_\theta C_1-2c_\alpha s_\alpha^3 c_\theta^2s_\theta C_3)
\ee
where $R$ is the AdS$_5$ radius and we abbreviate $s_\alpha=\sin\alpha,c_\alpha=\cos\alpha,s_\theta=\sin\theta,c_\theta=\cos\theta$. 

The dilaton is again given in terms of this $G$ as
\be
e^{2\Phi}=G\;.
\ee
Using $\mu,\nu$ for the five-dimensional (internal) coordinates, the metric components are given by:
\be
g_{\mu\nu}=G ~ \gt_{\mu\nu}
\ee
where 
\be
\begin{split}
\gt_{\alpha\alpha}&=R^2+\frac{w^2R^6}{2}\left[1-2c_\alpha^2+2c_\alpha^2s_\alpha^2c_\theta^2s_\theta^2 
+2c_\alpha^4+(4c_\alpha^2-3)c_\alpha s_\alpha  c_\theta s_\theta^2 C_2 -c_\alpha s_\alpha c_\theta^3 C_{3,1} \right. \\
&\qquad\qquad \qquad \;\;\left. 
-c_\alpha s_\alpha s_\theta^3 C_{2,1}+(4c_\alpha^4\!-\!3 c_\alpha^2\!+\!1) c_\theta s_\theta C_1
\!+\!(4c_\alpha^3-3 c_\alpha) s_\alpha c_\theta^2 s_\theta C_3\right]\;,
\end{split}
\ee

\be
\begin{split}
\gt_{\alpha\theta}&=\frac{w^2 R^6}2 \left[c_\alpha s_\alpha^3 c_\theta s_\theta (c_\theta^2-s_\theta^2)+s_\alpha^2 c_\alpha^2 s_\theta c_\theta^2 C_{3,1}
+(2 s_\alpha^2 s_\theta^2-s_\theta^2-s_\alpha^2) s_\alpha^2 c_\theta C_3\right.\\
 &\left.  \quad\quad\qquad -s_\alpha^2 c_\alpha^2 c_\theta s_\theta^2 C_{2,1}+c_\alpha^3 s_\alpha (1-2 s_\theta^2) C_1
+(2 s_\alpha^4 s_\theta^3-s_\alpha^2 s_\theta^3+s_\theta s_\alpha^2 c_\alpha^2) C_2\right]\;,
\end{split}
\ee

\be
\begin{split}
\gt_{\theta\theta}&=R^2\Big[s_\alpha^2+\frac{w^2R^4}2\left[s_\alpha^2-2s_\alpha^4s_\theta^2+
2s_\alpha^4 s_\theta^4+c_\alpha s_\alpha^3s_\theta(s_\alpha^2s_\theta^2+s_\theta^2 -s_\alpha^2)C_3
-c_\alpha^3 s_\alpha^3 s_\theta c_\theta^3 C_{2,1}
\right.\\
&\left. \quad\quad -c_\alpha^3 s_\alpha^3 c_\theta s_\theta^2 C_{3,1}
-c_\alpha s_\alpha^3c_\theta (s_\theta^2s_\alpha^2 -c_\theta^2 )C_2
+(s_\alpha^6+s_\alpha^4-2 s_\alpha^2) c_\theta s_\theta C_1 -s_\alpha^6 c_\theta s_\theta C_{3,2}
 \right]\Big] \;,
\end{split}
\ee

\be
\gt_{\alpha\phi_1}=\frac{w^2R^6}2\left[c_\alpha^2 s_\alpha^2 s_\theta c_\theta^2 S_3
+c_\alpha^2s_\alpha^2 s_\theta^3 S_{2,1}\!+\!c_\alpha(1+c_\alpha^2)s_\alpha c_\theta s_\theta S_1 +c_\alpha^2 s_\alpha^2 c_\theta^3 S_{3,1}+c_\alpha^2 s_\alpha^2c_\theta s_\theta^2 S_2\right]\;,
\ee

\be
\begin{split}
\gt_{\alpha\phi_2}&=\frac{w^2R^6}2\left[-s_\alpha^4 c_\theta s_\theta^4 S_2-c_\alpha s_\alpha^3 c_\theta s_\theta^3 S_1
+s_\alpha^2c_\alpha^2 s_\theta^3 S_{2,1}-s_\alpha^2 c_\alpha^2 c_\theta s_\theta^2 S_2+c_\alpha^3s_\alpha c_\theta s_\theta S_1
-s_\alpha^4 c_\theta^4 s_\theta S_3\right]\;,
\end{split}
\ee

\be
\begin{split}
\gt_{\alpha\phi_3}&=\frac{w^2R^6}2\left[ -s_\alpha^4 c_\theta s_\theta^4 S_2
-s_\alpha^4c_\theta^4s_\theta S_3-c_\alpha s_\alpha^3c_\theta^3s_\theta S_1
-s_\alpha^2c_\alpha^2c_\theta^2 s_\theta S_3 +c_\alpha^3 s_\alpha c_\theta s_\theta S_1
+s_\alpha^2c_\alpha^2c_\theta^3 S_{3,1}
\right]\;,
\end{split}
\ee

\be
\begin{split}
\gt_{\theta\phi_1}&=\frac{w^2 R^6} 2 \left[c_\alpha^3 s_\alpha^3 s_\theta^3 S_{3,1}+c_\alpha s_\alpha^3 s_\theta (c_\alpha^2 c_\theta^2-s_\alpha^2 s_\theta^2) S_2
-c_\alpha s_\alpha^3 c_\theta s_\theta^2 S_3+c_\alpha^3 s_\alpha^3 c_\theta s_\theta^2 S_{2,1}\right. \\
&\left.\qquad\qquad \quad-2 c_\alpha^4 s_\alpha^2 s_\theta^2 S_1  -c_\alpha^3 s_\alpha^3 s_\theta S_{3,1}+c_\alpha s_\alpha^5 c_\theta S_3+c_\alpha^4 s_\alpha^2 S_1\right]\;,
\end{split}
\ee

\be
\begin{split}
\gt_{\theta\phi_2}&=\frac{w^2 R^6}2\left[-c_\alpha s_\alpha^3 s_\theta (2 s_\alpha^2 s_\theta^2+c_\theta^2-s_\alpha^2 s_\theta^4) S_2
-s_\alpha^2 s_\theta^2 (c_\alpha^2-s_\alpha^2 s_\theta^2 c_\alpha^2) S_1\right.\\
&\left.\qquad\qquad\quad  -c_\alpha s_\alpha^5 c_\theta s_\theta^4 S_3+c_\alpha^3 s_\alpha^3 c_\theta s_\theta^2 S_{2,1}
-s_\alpha^6 s_\theta^2 c_\theta^2 S_{3,2}\right]\;,
\end{split}
\ee

\be
\begin{split}
\gt_{\theta\phi_3}&=\frac{w^2R^6}2\left[c_\alpha s_\alpha^5c_\theta^4 s_\theta S_2-c_\alpha^3 s_\alpha^3c_\theta^2 s_\theta S_{3,1}
-c_\alpha s_\alpha^5 c_\theta^5 S_3-s_\alpha^4c_\alpha^2 c_\theta^4 S_1-s_\alpha^6c_\theta^2 s_\theta^2 S_{3,2}\right.\\
&\left.\qquad \qquad -c_\alpha s_\alpha^3(1-2s_\alpha^2)c_\theta^3 S_3+s_\alpha^2c_\alpha^2 c_\theta^2 S_1
+c_\alpha s_\alpha^3c_\theta S_3\right]\;,
\end{split}
\ee

\be
\begin{split}
\gt_{\phi_1\phi_1}&=R^2\Big[c_\alpha^2+\frac{w^2R^4}2\left[-c_\alpha^3 s_\alpha^3 c_\theta^2 s_\theta C_3+c_\alpha^3s_\alpha^3 s_\theta^3 C_{2,1}
-(c_\alpha^2 s_\alpha^4+6 c_\alpha^4 s_\alpha^2) c_\theta s_\theta C_1  \right. \\
& \left.\qquad\qquad
+c_\alpha^3 s_\alpha^3 c_\theta^3 C_{3,1}-c_\alpha^3 s_\alpha^3 c_\theta s_\theta^2 C_2+c_\alpha^2 s_\alpha^4+c_\alpha^4 s_\alpha^2+2c_\alpha^6\right]\Big]\;,
\end{split}
\ee

\be
\begin{split}
\gt_{\phi_1\phi_2}&=-\frac{w^2 R^6}2 \left[c_\alpha^4 s_\alpha^2 s_\theta^2+c_\alpha^2 s_\alpha^4 s_\theta^4-2 c_\alpha^2 s_\alpha^4 c_\theta^2 s_\theta^2
+(c_\alpha s_\alpha^5 c_\theta s_\theta^4-c_\alpha^3 s_\alpha^3 s_\theta^2 c_\theta) C_2\right. \\
&\left.\qquad\qquad\quad +(c_\alpha^4 s_\alpha^2 c_\theta s_\theta-c_\alpha^2 s_\alpha^4 c_\theta s_\theta^3) C_1  -c_\alpha^3 s_\alpha^3 s_\theta^3 C_{2,1}+c_\alpha s_\alpha^5 c_\theta^4 s_\theta C_3\right]\;,
\end{split}
\ee

\be
\begin{split}
\gt_{\phi_1\phi_3}&=-\frac{w^2R^6}2\left[c_\alpha s_\alpha^5 c_\theta s_\theta^4 C_2-2 c_\alpha^2 s_\alpha^4 c_\theta^2 s_\theta^2
+c_\alpha s_\alpha^5 c_\theta^4 s_\theta C_3 -c_\alpha^2 s_\alpha^4 c_\theta^3 s_\theta C_1 -c_\alpha^3 s_\alpha^3 c_\theta^2 s_\theta C_3\right.\\
&\left.\qquad\qquad\quad 
+c_\alpha^4 s_\alpha^2c_\theta s_\theta C_1 +c_\alpha^2 s_\alpha^4 c_\theta^4-c_\alpha^3 s_\alpha^3 c_\theta^3 C_{3,1}
+c_\alpha^4 s_\alpha^2 c_\theta^2\right]\;,
\end{split}
\ee

\be
\begin{split}
\gt_{\phi_2\phi_2}&=R^2\Big[s_\alpha^2s_\theta^2+\frac{w^2R^4}2\left[s_\alpha^2s_\theta^2(1-s_\alpha^2 s_\theta^2+2s_\alpha^4s_\theta^4)
+s_\alpha^6c_\theta^3s_\theta^3 C_{3,2} -c_\alpha s_\alpha^5 s_\theta^3c_\theta^2 C_3\right.\\
&\left.\qquad \qquad \quad -c_\alpha^2s_\alpha^4c_\theta s_\theta^3 C_1
+c_\alpha^3 s_\alpha^3 s_\theta^3 C_{2,1}
-c_\alpha s_\alpha^3c_\theta s_\theta^2 (6s_\alpha^2s_\theta^2 + s_\alpha^2 c_\theta^2
+c_\alpha^2) C_2\right]\Big]\;,
\end{split}
\ee

\be
\begin{split} 
\gt_{\phi_2\phi_3}&=\frac{w^2 R^6}2 \left[2 c_\alpha^2 s_\alpha^4 c_\theta^2 s_\theta^2-s_\alpha^6 c_\theta^2 s_\theta^2+c_\alpha s_\alpha^5 c_\theta s_\theta^2 (c_\theta^2-s_\theta^2) C_2
+s_\alpha^6 c_\theta^3 s_\theta^3 C_{3,2}\right.\\
&\left.\qquad\qquad+c_\alpha s_\alpha^5 c_\theta^2 s_\theta (s_\theta^2-c_\theta^2) C_3-c_\alpha^4 s_\alpha^2 c_\theta s_\theta C_1\right]\;,
\end{split}
\ee

\be
\begin{split}
\gt_{\phi_3\phi_3}&=R^2\Big[s_\alpha^2c_\theta^2+\frac{w^2R^4}2\left[s_\alpha^2 c_\theta^2(1-s_\alpha^2 c_\theta^2 +2 s_\alpha^4 c_\theta^4)
+s_\alpha^6 c_\theta^3 s_\theta^3 C_{3,2}-s_\alpha^4 c_\alpha^2 c_\theta^3s_\theta C_1\right.\\ 
&\left.\qquad\qquad \;\; -5 c_\alpha s_\alpha^5 c_\theta^4s_\theta C_3-c_\alpha s_\alpha^3 c_\theta^2 s_\theta C_3-c_\alpha s_\alpha^5 c_\theta^3s_\theta^2 C_2
+c_\alpha^3 s_\alpha^3 c_\theta^3 C_{3,1}\right]\Big]\;.
\end{split}
\ee
When transforming to spherical coordinates, the $B$-field also transforms to a purely five-dimensional
field, as of course required for the solution to be a direct product. Its components are:
\be
\begin{split}
B_{\alpha\theta}&=w R^4 G \left(s_\alpha^2s_\theta S_2+s_\alpha^2c_\theta S_3+c_\alpha s_\alpha S_1\right)/2\;,\\
B_{\alpha\phi_1}&=w R^4 G \left(2c_\alpha s_\alpha s_\theta^2-c_\alpha s_\alpha\right)/2\;,\\
B_{\alpha\phi_2}&=w R^4 G \left(s_\alpha^2 c_\theta s_\theta^2 C_2+c_\alpha s_\alpha s_\theta^2-s_\alpha^2 c_\theta^2 s_\theta C_3
-c_\alpha s_\alpha c_\theta s_\theta C_1\right)/2\;,\\
B_{\alpha\phi_3}&=w R^4 G \left(s_\alpha^2c_\theta s_\theta^2 C_2+c_\alpha s_\alpha s_\theta c_\theta C_1-s_\alpha^2c_\theta^2s_\theta C_3
-c_\alpha s_\alpha c_\theta^2\right)/2\;,\\
B_{\theta\phi_1}&=w R^4 G \left(2c_\alpha^2s_\alpha^2c_\theta s_\theta-c_\alpha s_\alpha^3s_\theta C_2-c_\alpha s_\alpha^3 c_\theta C_3+c_\alpha^2 s_\alpha^2 c_1\right)/2\;,\\
B_{\theta\phi_2}&=w R^4 G\left(-c_\alpha s_\alpha^3 s_\theta^3 C_2+c_\alpha s_\alpha^3 c_\theta s_\theta^2 C_3+s_\alpha^2c_\alpha^2 s_\theta^2 C_1+s_\alpha^2(1-2s_\alpha^2) c_\theta s_\theta\right)/2\;,\\
B_{\theta\phi_3}&=wR^4 G \left(c_\alpha s_\alpha^3 c_\theta^2 s_\theta C_2+s_\alpha^2(1-2 s_\alpha^2) c_\theta s_\theta
-c_\alpha s_\alpha^3 c_\theta^3 C_3+s_\alpha^2 c_\alpha^2 c_\theta^2 C_1\right)/2\;,\\
B_{\phi_1\phi_2}&=wR^4 G\left(-c_\alpha s_\alpha^3 c_\theta s_\theta^2 S_2+c_\alpha s_\alpha^3c_\theta^2 s_\theta S_3-c_\alpha^2s_\alpha^2 c_\theta s_\theta S_1\right)/2\;,\\
B_{\phi_1\phi_3}&=wR^4 G\left(-c_\alpha s_\alpha^3 c_\theta s_\theta^2 S_2+c_\alpha s_\alpha^3c_\theta^2 s_\theta S_3+c_\alpha^2s_\alpha^2 c_\theta s_\theta S_1\right)/2\;,\\
B_{\phi_2\phi_3}&=wR^4 G\left(-c_\alpha s_\alpha^3 c_\theta s_\theta^2 S_2-c_\alpha s_\alpha^3c_\theta^2 s_\theta S_3+c_\alpha^2s_\alpha^2 c_\theta s_\theta S_1\right)/2\;.\\
\end{split}
\ee
Let us also note the determinant of the metric:
\be
\sqrt{\text{det}g_w}=R^5 G \sin^3\alpha\cos\alpha\sin\theta\cos\theta\;,
\ee
from which we can construct the Hodge dual of $H_{\mu\nu\rho}$,
\be
 \tilde{H}_{\mu\nu}=\frac{1}{24}\sqrt{\text{det}g_w}\epsilon_{\mu\nu\rho\sigma\tau} H^{\rho\sigma\tau}
\ee
and thus the three-form field strength $F^{(3)}=d C^{(2)}$ of the RR two-form:
\be
F^{(3)}=\diff( e^{-2\Phi} \tilde{H})\;.
\ee
The expression of $F_{\mu\nu\rho}$ is too long to be reproduced here but it can be easily constructed
from the ancillary files provided with the submission. 

The five-form field strength $F_5=d C_4$ is taken to be of the same form as the other $AdS_5\times \text{sphere}$ metrics,
with the $w$-deformed geometry reflected only in the coefficient $G$ in front of the internal metric:
\be
F=R^5(\omega_{\text{AdS}_5}+G \omega_{\Srm^5})\;,\quad\text{with} \quad \omega_{\Srm^5}=s_\alpha^3c_\alpha s_\theta c_\theta \diff \alpha\wedge \diff\theta\wedge \diff \phi_1\wedge\diff\phi_2\wedge\diff\phi_3
\ee
Finally the RR zero-form $C_0$ is taken to vanish, as in \cite{Kulaxizi:2006zc} (and as required by consistency with the 
forms of the metric and five-form). 

We have succeeded in writing down the full R-R geometry of the $w$-deformation by making use of the gauge-theoretic 
noncommutative structure. As expected, the $\Urm(1)\times \Urm(1)$ isometries are not directly visible in the coordinate
system above, which is the most natural one for the $w$-deformed gauge theory. The unfamiliar form can of
course be brought into the form of the LM solution \cite{Lunin:2005jy} by the coordinate transformation 
following from (\ref{zredefinition}).

\subsection{Visualising the deformations}

\begin{figure}[h]
\begin{center}
\hspace{-1cm}(a)~\includegraphics[width=4cm]{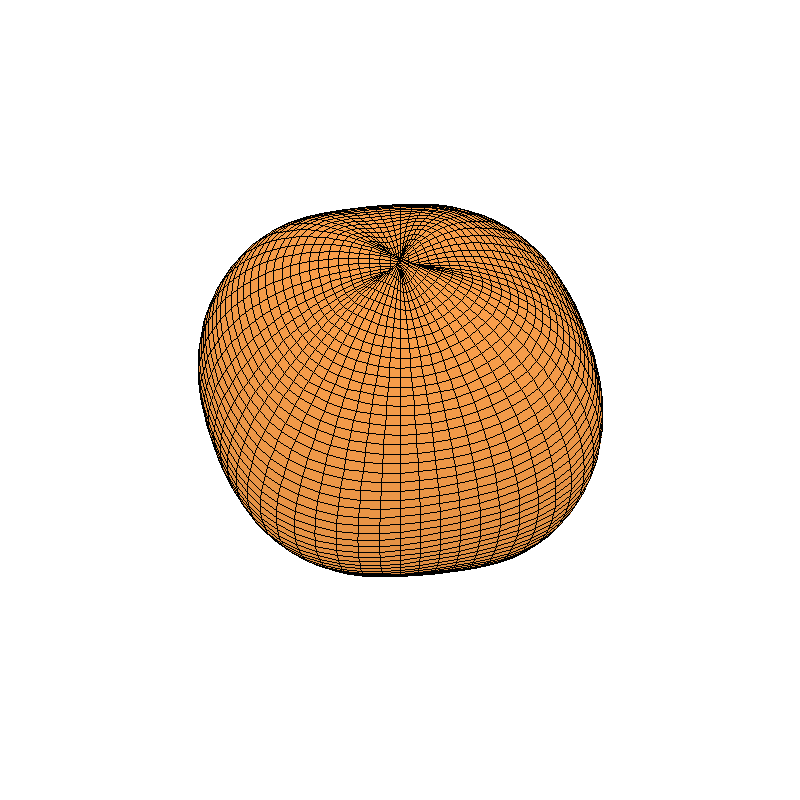}{\hspace{-2.6cm}$\beta R$=0.3\hspace{1cm}}
\includegraphics[width=4cm]{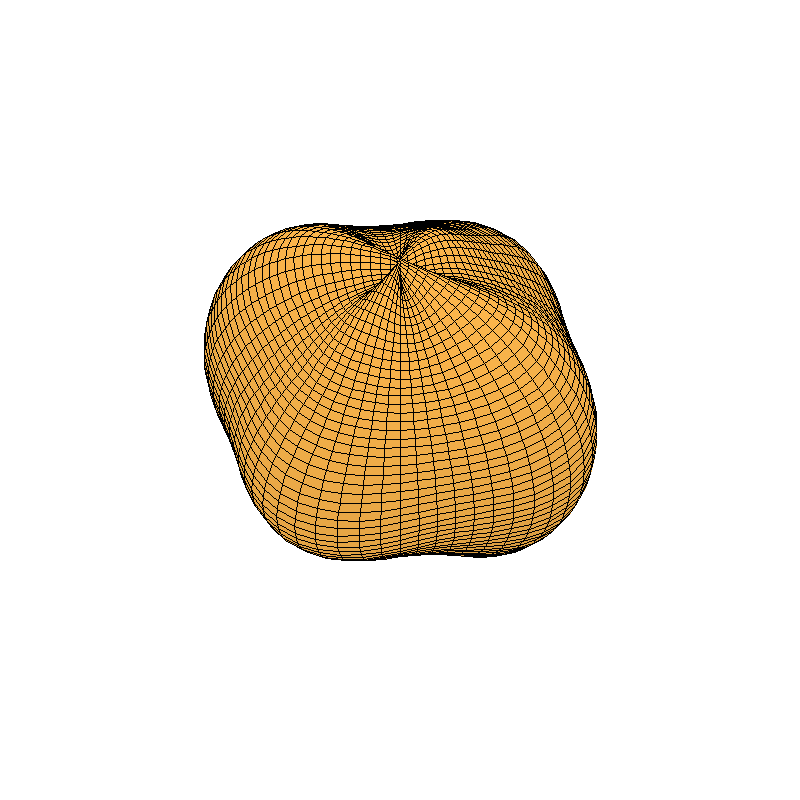}{\hspace{-2.7cm}$\beta R$=0.45\hspace{1cm}}
\includegraphics[width=4cm]{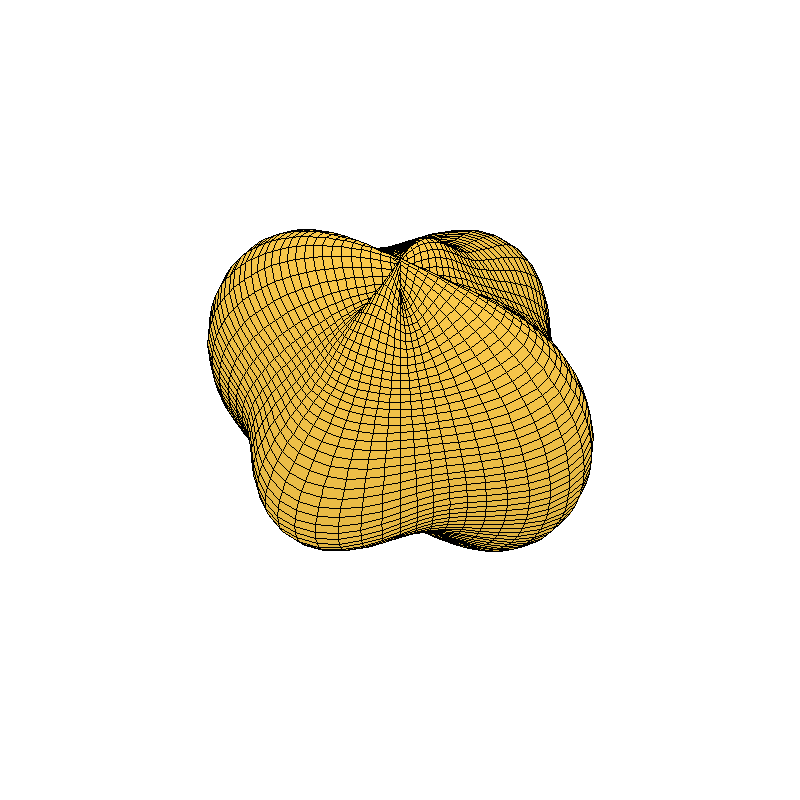}{\hspace{-2.6cm}$\beta R$=0.6\hspace{1cm}}
\includegraphics[width=4cm]{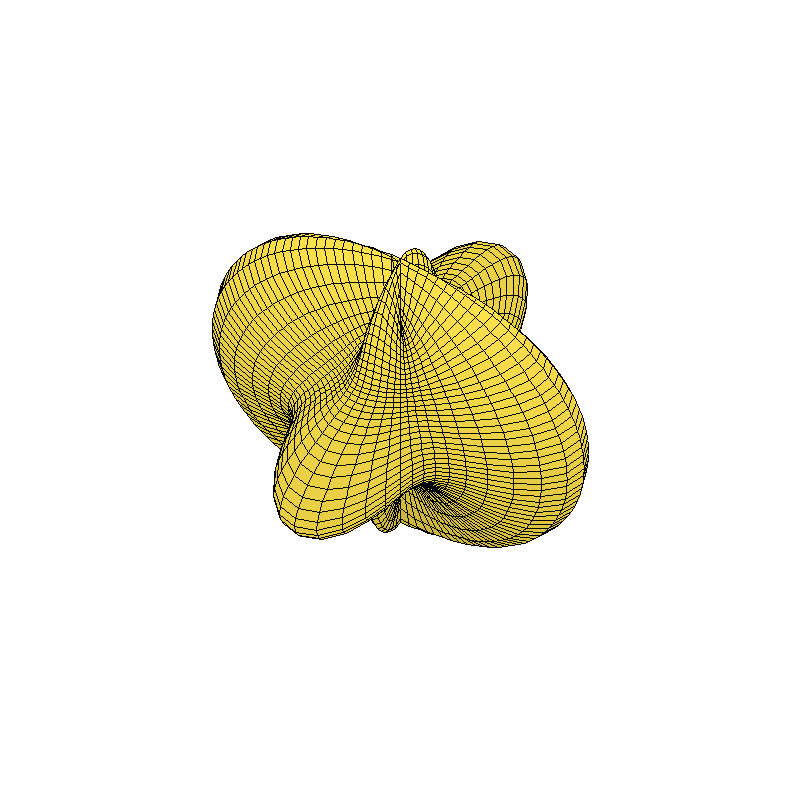}{\hspace{-2.6cm}$\beta R$=1.0}\\
\hspace{-1cm}(b)~\includegraphics[width=4cm]{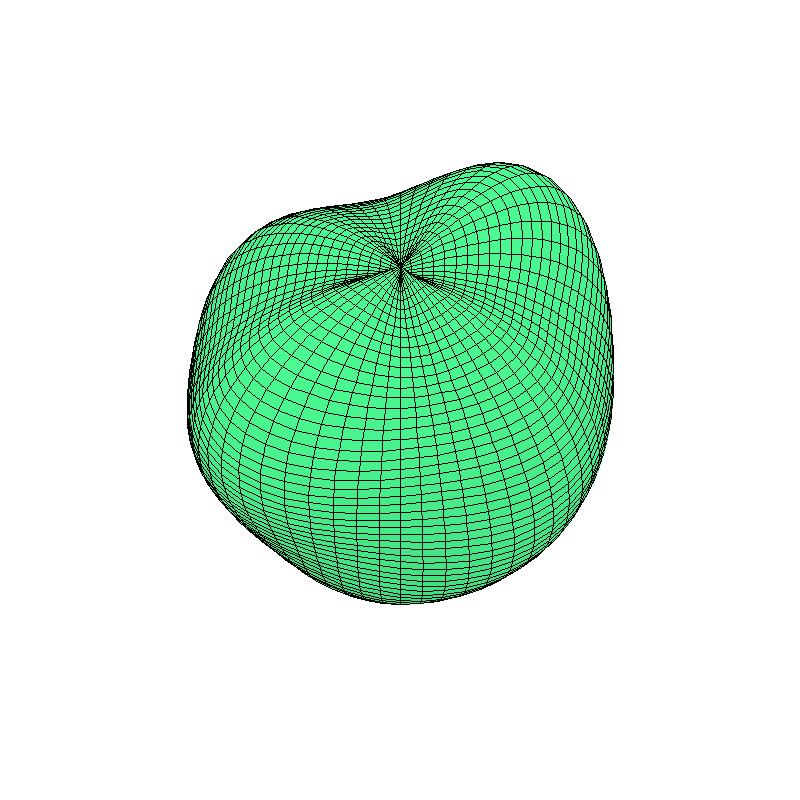}{\hspace{-2.6cm}$wR$=0.3\hspace{1cm}}
\includegraphics[width=4cm]{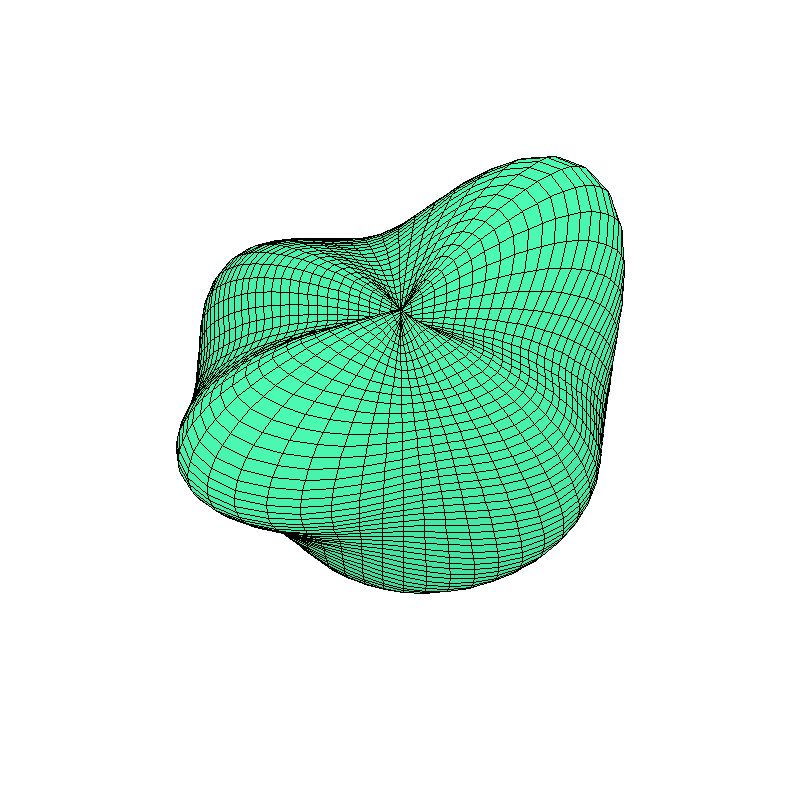}{\hspace{-2.7cm}$wR$=0.45\hspace{1cm}}
\includegraphics[width=4cm]{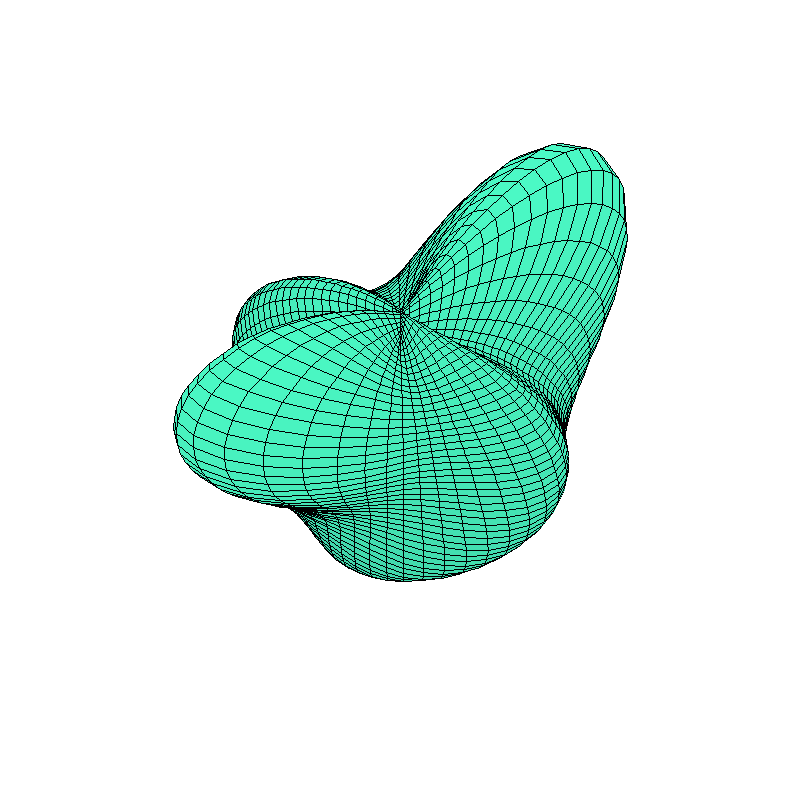}{\hspace{-2.6cm}$wR$=0.6\hspace{1cm}}
\includegraphics[width=4cm]{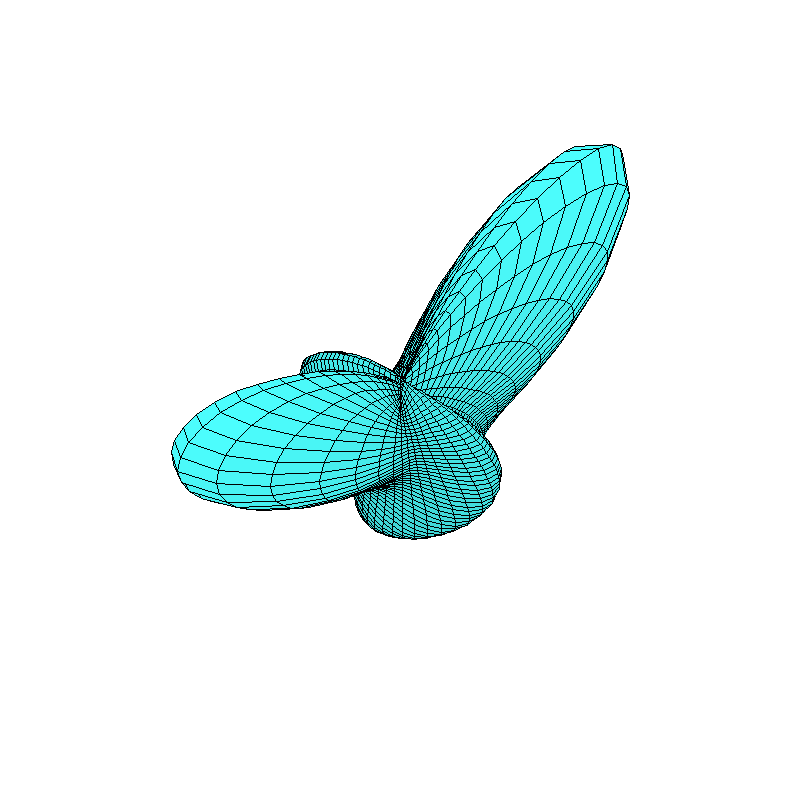}{\hspace{-2.6cm}$wR$=1.0}
\end{center}

\caption{Plot of the $\alpha,\theta$ dependence of the scalar curvature of the (a) $\beta$-deformed and (b) 
$w$-deformed metrics as a function of their respective
parameters. For the $w$-deformation we have restricted to the subspace $\phi_i=0$. We have suppressed the axes
for clarity. The distance from the
origin corresponds to the magnitude of the scalar curvature, which in the undeformed case is simply $20/R^2$. 
However the 
scale changes from plot to plot so they are not directly comparable. We emphasise that the plots are \emph{not}
$\text{R}^3$ embeddings of the geometries.} 
\label{wdefplot}
\end{figure}

 Although the metric of the $w$-deformed theory above is equivalent to that of the $\beta$-deformation, 
it is written in a coordinate system that is more adapted to computations with the $w$-deformed theory.
 A quick way of seeing the difference between the coordinate systems is by plotting the scalar curvature 
as a function of the $\alpha,\theta$ angles, and as a function of the deformation parameter. This is done 
in Fig. \ref{wdefplot} for the slice where we take the three $\phi_i$ angles to zero. The $\alpha,\theta$ angles
 are taken to be the polar and azimuthal angles of a usual spherical coordinate system. The apparent difference
in the scalar curvatures is due to the fact that the $\alpha,\theta$ angles are not directly
comparable in the two coordinate systems (in particular, in the $w$-frame the scalar curvature depends
on the $\phi_i$ angles as well as on $\alpha,\theta$), as well as the relation $\beta^2=3w^2$. 

Of course, ideally, to better visualise the geometry of a higher-dimensional deformed sphere one would like to 
construct an isometric embedding in $\text{R}^3$. For the
imaginary-$\beta$ deformed LM sphere, such an embedding was recently provided in \cite{Jagrell14}.

\section{Outlook}

In this section we touch upon various potential extensions and applications of our work.

\paragraph{Relation to other approaches} \mbox{}

An important recent development in understanding the algebraic structures behind AdS/CFT has
been the gravity/classical YBE correspondence \cite{Delduc:2013qra,Hoare:2014pna,Kawaguchi:2014qwa,Kawaguchi:2014fca,
Matsumoto:2014nra,Matsumoto:2014gwa,vanTongeren:2015soa,vanTongeren:2015uha},
where the deformation is performed at the level of the string sigma model and can be used as a supergravity
generating technique \cite{Lunin:2014tsa,Hoare:2015wia}. Similarly to our method, these deformations are based
on a classical $r$-matrix, which satisfies an, in general, modified classical YBE. In the unmodified CYBE case
this approach generates IIB supergravity solutions (and indeed the LM solution has been reproduced in this 
framework in \cite{Matsumoto:2014nra}) while in the modified CYBE case it has recently been shown to produce solutions 
of a modified system of equations which are related to standard IIB solutions by T-duality 
\cite{Arutyunov:2015qva,Arutyunov:2015mqj}. It would certainly be interesting to explore possible links with our approach. 

A parallel line of work has been the application of non-abelian T-duality to interpolate between CFT's while 
preserving integrability \cite{Sfetsos:2013wia,Sfetsos:2010uq}. These $\lambda$-deformations
have been shown to be closely related to the $\eta$ deformations above \cite{Hoare:2015gda,Sfetsos:2015nya,Klimcik:2015gba}. 
A recent review can be found in \cite{Thompson:2015lzd}.
 
A further way to deform the $\Ncal=4$ SYM integrable structure is at the level of the centrally extended 
$\text{psu}(2|2)\otimes \text{psu}(2|2)$ symmetry algebra of the integrable S-matrix of the theory,
as in the work \cite{Beisert:2008tw}. For real deformation parameter $q$, these 
deformations have been related to the $\eta$ deformations above \cite{Arutyunov:2013ega}, while
for $q$ a root of unity they have been studied in \cite{Hollowood:2014qma}.

The general feature of all the sigma model-based deformations above is that they explicitly preserve integrability\footnote{Although
they can also be applied to non-integrable backgrounds \cite{Crichigno:2014ipa}, so integrability does not appear to be a 
prerequisite.}, but tend to lack a 
clear gauge theory interpretation. Our approach is in principle complementary in that it is motivated by a well-understood
gauge theory deformation, but the sigma-model side is less clear, and the twist of the supergravity structures,
although well motivated, does not follow directly from the gauge theory twist. So perhaps a method including aspects
of the above approaches could lead to a more powerful solution-generating technique that extends beyond the 
Leigh-Strassler deformations. As a first step, one would like to understand whether our deformation can be 
performed at the level of the string sigma model, which would provide additional insight on integrability 
(and its absence). 

We have frequently remarked on the similarities between our approach and that of \cite{Kulaxizi:2006zc},
which used the Seiberg-Witten relations \cite{Seiberg:1999vs} to relate a noncommutative deformation on 
the gauge theory side to a deformed gravity background. It appears likely that, at least for integrable
deformations, our approach is actually
equivalent to \cite{Kulaxizi:2006zc}, and it would be interesting to show this explicitly. It is less
clear whether the approaches would still be equivalent in the non-integrable cases. If this were to be
true, a possible extension of our methods to the non-associative case (see below) might lead to 
an extension of Seiberg-Witten-type methods to the non-associative case as well.

\paragraph{Extension to the complex-$w$ background} \mbox{}

The work \cite{Kulaxizi:2006zc} constructed two different matrices for the $\rho$ (pure-$h$) deformation, which were 
associated to the real and imaginary part of the $\rho$ parameter respectively. Since (\ref{wThetamatrix}) led
to the real-$w$ deformation, we would similarly expect that there 
will be a noncommutativity matrix for the case where $w$ is imaginary, and indeed a candidate such matrix is
\be \label{wimaginary}
\Theta^{IJ}=\begin{pmatrix}0&z_1 z_2-z_3^2&z_2^2-z_1z_3& 0&-X& \bar{X}\\
z_3^2-z_1z_2& 0& z_2z_3-z_1^2& \bar{X}& 0& -X \\
z_1z_3-z_2^2& z_1^2-z_2z_3& 0& -X& \bar{X}& 0\\
0& -\bar{X}& X&0& \zb_1\zb_2-\zb_3^2& \zb_2^2-\zb_3\zb_1\\
X& 0& -\bar{X}& \zb_3^2-\zb_1\zb_2& 0& \zb_2\zb_3-\zb_1^2\\
-\bar{X}&X& 0& \zb_3\zb_1-\zb_2^2& \zb_1^2-\zb_2\zb_3& 0
 \end{pmatrix}
\ee
where $X=z_1\zb_2+z_2\zb_3+z_3\zb_1$. This matrix is similar to the one found to correspond to the imaginary-$h$
deformation in \cite{Kulaxizi:2006zc}. 

This matrix also converts to an $r$-independent deformation matrix in spherical coordinates and is thus a 
good candidate to produce a conformal theory. It is expected to correspond to the non-integrable version of 
the $w$-twist when the $w$ parameter becomes complex. The non-integrability appears to be reflected in that the 
pure spinor pair derived from (\ref{wimaginary}) fails to be compatible at first order and higher-order 
corrections in $w$ need to be added. Understanding the role of associativity in deriving these corrections 
is of crucial importance, and we aim to report on a fuller analysis of this case in the near future.\footnote{An
intermediate step would be to construct the complex $\beta$-deformation of \cite{Lunin:2005jy} 
using an extension of our techniques, which is work currently in progress.}

An alternative way to construct the complex-$w$-deformed background might proceed through  IIB S-duality from
the real-$w$ solution, as in  \cite{Lunin:2005jy}. Since the relation (\ref{Tmatrix}) between the 
$\beta$- and $w$-deformation applies only for real values of the parameters, it will be interesting to determine
whether the resulting background is the same as the complex-$\beta$ background in \cite{Lunin:2005jy}. 

Assuming that the complex-$w$ background is indeed different to the complex-$\beta$ one, its
construction would be important, since it 
would, for instance, allow detailed studies of integrable vs. non-integrable string motion along the 
lines of \cite{Giataganas:2013dha} for the real vs. complex $\beta$-deformation.

\mparagraph{Extension to the generic Leigh-Strassler theory}

The triangularity condition (\ref{triangular}) satisfied by the $R_{q,h}$ matrix is a strong indication
that it is possible to construct a twist relating that matrix to the undeformed $\Ncal=4$ SYM $R$-matrix.
The close resemblance of the noncommutativity matrices of the $w$
deformation to a linear combination of $\beta$ and $h$ noncommutativity matrices (as derived in 
\cite{Kulaxizi:2006zc}) can be taken as further evidence for the existence of a generic twist. 
The anticipated problem, however, in applying our 
methods beyond the integrable Leigh-Strassler theories is not so much the construction of a suitable twist 
but dealing with the non-quasitriangular structures it is expected to lead to.  

In this work we have focused on Hopf twists, which by definition do not alter the quasitriangular
structure of the theory. So, starting with the $\Ncal=4$ SYM structure, they can only be applied to obtain
other integrable cases. A promising way to twist from $\Ncal=4$ SYM to a non-integrable Leigh-Strassler
theory would be to make use of \emph{quasi-Hopf twists} \cite{Drinfeld90} (see
\cite{Mylonas:2013jha} for a recent discussion of such twists in a related context). These twists
lead to non-associative quasi-Hopf algebras. If it can be shown that the
non-integrable Leigh-Strassler deformations  arise in this way, the question would then be 
whether and how one can translate a quasi-Hopf noncommutative structure to the gravity side in order
to produce the dual background. Work in this direction is in progress.

\mparagraph{Other extensions} \mbox{}

The deformations we considered involve only the $\SU(3)$ part of the $\SU(4)\sim \SO(6)$ global symmetry
of the $\Ncal=4$ theory. The reason for this is the wish to preserve supersymmetry. One can certainly try
to explore deformations involving the full $\SU(4)$, which would not be supersymmetric but can be expected to lead
to planar-integrable subsectors similar to the $\gamma_i$ deformations \cite{Frolov:2005dj}. 

One can also contemplate an extension of our noncommutativity-based techniques to conformal theories
beyond the Leigh-Strassler deformations, such
as the $\Ncal=2$ SQCD whose integrability properties have been extensively studied starting 
with \cite{Pomoni:2011jj,Gadde2012a,Pomoni:2013poa}. 

It should be emphasised that, even  though our construction is purely in the planar limit (in the sense
of relying on planar structures like the R-matrix which is derived from the spin-chain Hamiltonian), 
the $\widetilde{\SU(3)}_{q,h}$ symmetry derived in \cite{Mansson:2008xv} and used here is a symmetry of 
the Leigh-Strassler lagrangian and does appear to require the planar limit. On the other hand, the
conformality constraint on the Leigh-Strassler couplings $g,\kappa,q,h$ does depend on the rank
$N$ of the gauge group \cite{Aharony:2002tp,Freedman:2005cg,Penati:2005hp,Rossi:2005mr}. It is thus important 
to understand how adding $1/N$ corrections might affect the algebraic structures we have discussed.  

Another important issue concerns the role of the spectral parameter. Recall that all our discussion
has been in the limit where the spectral parameter is taken to infinity. It is worth considering whether the
spectral parameter can be put back, at least for the integrable cases, and what its interpretation
would be on the gauge theory side. Also, since our method is based on structures (such as the $R$-matrix)
appearing at one loop
in the gauge theory, understanding the (possible) extension of these structures to higher loops would
provide additional insight and confidence in their application.

\section{Conclusion}

In this work, we proposed a way of making the Hopf-algebraic symmetry of a class of conformal gauge
theories visible on the gravity side of the AdS/CFT correspondence. 
The main elements of our construction were:
\begin{itemize}
\item The underlying Hopf algebra/noncommutative structure of the gauge theory
\item The generalised geometry description of $\Ncal=2$ IIB solutions
\item Applying noncommutativity to construct the deformed pure spinors of the background. 
\end{itemize}

We saw that these steps led to the construction of the dual supergravity background of the $\beta$-deformation
and of its equivalent $w$-deformation. Although  the $w$-deformed solution can also be constructed 
directly from the $\beta$-deformed one, its direct derivation from $\Ncal=4$ SYM presented above  is particularly 
interesting since it exhibits more of the features that one might expect to encounter in attempting to construct the
duals of more general deformations. In particular, the natural coordinate system for the $w$-deformation obscures
the isometries of the metric, making the verification of the solution computationally challenging. This is also
expected to be true (and, likely, exacerbated) for more  general deformations.

One of our main motivations in this work was to clarify the role of the Hopf algebra symmetry of \cite{Mansson:2008xv}
in the Leigh-Strassler theories (and perhaps eventually in other finite four-dimensional theories). Does the
symmetry extend beyond one-loop? Can it be used to classify observables like any other global symmetry? The fact 
that we have found an analogue of this symmetry on the gravity side is an encouraging sign that the symmetry is 
more than a one-loop artifact and that (at least in the integrable cases) it plays an important role also on 
the strong coupling side. More studies (also from the string-sigma model side, which we have not considered 
in the current work) are needed to fully understand the consequences of this intriguing quantum group symmetry. 

Of course, the quest to construct the AdS/CFT dual of the generic Leigh-Strassler deformation continues. We hope 
that our techniques will provide useful new input in that direction. 

\mbox{}

\mparagraph{Acknowledgements}

KZ is thankful to Manuela Kulaxizi for very useful discussions and comments on the manuscript,
Teresia M\aa nsson for discussions and collaboration during the early stages of this project
and Robert de Mello Koch for discussions. We are also grateful to Stijn van Tongeren for very helpful 
comments on the first version of the manuscript. KZ is supported by the South African National Research
Foundation (NRF) under grant CSUR 93735. HD is supported by the South African National Institute
for Theoretical Physics (NITheP) through an MSc bursary.

\appendix

\section{Hopf algebras} \label{Hopfappendix}

In this appendix we introduce some basic properties of Hopf algebras in order to fix notation and
for quick reference. For a more thorough exposition, the reader is referred to any of the many
excellent books on quantum books, for instance \cite{Majid}. Note that when working with elements belonging
to two copies of a vector space $V$ we will often use Sweedler notation, in which an element 
$F^{i\;\;j}_{\;k\;\;l}=\sum_a (F^{a(1)})^i_{\;k}\otimes (F^{a(2)})^j_{\;l}$ of $V\otimes V$
is expressed as
\be
F=\sum_a F_{a(1)}\otimes F_{a(2)}\quad \text{or more simply} \quad F=F_{(1)}\otimes F_{(2)}\;.
\ee
The condensed notation takes some getting used to but can greatly simplify expressions, especially
involving coproducts. 

 A Hopf algebra (often called a quantum group) can be thought of as a generalisation of a usual 
Lie algebra such that both the algebra product and co-product are non-commutative. Recall
that the product takes two elements of the algebra to a single element:
\be
m(C\otimes C)=C\cdot C\ra C\;.
\ee
In the universal enveloping algebra $U(g)$ of a (non-abelian) Lie algebra $g$, this product is matrix multiplication in
a given representation of the generators, and is thus non-commutative. The co-product, on the other hand, takes 
a single element to two elements:
\be
\Delta(C)\ra C\otimes C\;.
\ee
A familiar application of the coproduct is in the quantum mechanical addition of spins. When 
acting with the angular momentum operator $J$ on a product of two wavefunctions, we are really
first applying a coproduct, in the form
\be
\Delta(J)=J\otimes 1 +1\otimes J
\ee
We call this a \emph{trivial} action of the coproduct. It has the special property of being 
\emph{co-commutative}, meaning $\tau \circ \Delta (J) = \Delta(J)$, where 
$\tau: V_{1} \otimes V_{2} \rightarrow V_{2} \otimes V_{1}$ is the transposition map.

In Hopf algebras, such as the quantum group $su(2)_q$, the coproduct is deformed for some of the 
generators and becomes non-co-commutative. An interesting way to relax the co-commutativity axiom 
is so that it holds up to conjugation by an invertible element $R \in \Hcal \otimes \Hcal$. 
Such an element $R$ is called a \textit{quasitriangular structure}. Formally: A quasitriangular 
Hopf algebra is a pair $(\Hcal,R)$ where $\Hcal$ is a Hopf algebra and $R$ is an invertible element 
of $\Hcal \otimes \Hcal$ satisfying:
\be
\begin{split} \label{quasitriangularR}
(\Delta \otimes \id) R = &R_{13}R_{23}, \ \ (\id \otimes \Delta)R = R_{13} R_{12} \\
\tau \circ \Delta(h) &= R(\Delta(h))R^{-1}, \ \ \forall h \in \Hcal.
\end{split}
\ee
Here the indexed form of $R$ is to be understood as
$R_{ij} = \sum 1 \otimes \cdots \otimes R_{i} \otimes \cdots \otimes R_{j} \otimes 1 $ and the same
notation will be used for the twist $F$. We will often write simply $R$ instead of $R_{12}$. 

Given a Hopf algebra $\Hcal$, it is possible to obtain a new Hopf algebra, $\Hcal_{F}$, via the 
twisting construction. This requires
the existence of an invertible element $F \in \Hcal \otimes \Hcal$, called a 2-cocycle, such that 
\begin{equation} \label{cocyclecondition}
 (1 \otimes F)(\id \otimes \Delta)F = (F \otimes 1)(\Delta \otimes \id)F   
\end{equation}
A 2-cocycle satisfying this condition, as well as the counital relation 
$(\epsilon \otimes \text{id})F=1=(\text{id} \otimes \epsilon)F$
with $\epsilon$ the counit of the algebra, is also known as a \emph{Drinfeld} or \emph{Hopf twist}.\footnote{``Drinfeld twist'' is
the standard nomenclature. We use ``Hopf twist'' to emphasise the difference from twists leading to quasi-Hopf algebras, 
which we call ``quasi-Hopf twists''.}
The benefit of the twist construction is that it retains additional structures on the original Hopf algebra $\Hcal$. For instance,
 if $(\Hcal,R)$ be a quasitriangular HA with $R$ the quasitriangular structure, the twisted Hopf algebra 
($\Hcal_{F} \ ,\ R_{F}$) is also quasitriangular. The co-product, quasitriangular structure and antipode of the original algebra are all twisted to become
\begin{gather} \label{twistedstructures}
    \Delta_{F}(h)=F (\Delta (h)) F^{-1}  \ \forall \ h \in \Hcal \\
     R_{F} = F_{21}R F^{-1} \\
     S_{F}h = U(Sh)U^{-1} \ \forall \ h \in \Hcal
\end{gather}
with U invertible and given by $U=\sum F^{(1)}(S F^{(2)})$. The deformed coproduct  is
used in section \ref{StarProduct} to construct the deformed star product on the coordinates of 
$\Cset_3$, the module on which the Hopf algebra $\Hcal$ acts. 

It can be shown (see e.g. \cite{Majid} for a proof) that if (\ref{quasitriangularR}) hold then
$R$ satisfies the Yang-Baxter equation
\be
R_{12}R_{13}R_{23}=R_{23}R_{13}R_{12}
\ee
which, in components, takes the form
\be
R^{i_1 i_2}_{\;j_1 j_2} R^{j_1 i_3}_{\;k_1j_3} R^{j_2 j_3}_{\;k_2 k_3}=R^{i_2 i_3}_{\;j_2 j_3} R^{i_1 j_3}_{\;j_1 k_3} R^{j_1 j_2}_{\;k_1 k_2}\;.
\ee
We will thus also call $R$ the $R$-matrix of the Hopf algebra. In the context of Hopf algebras defined through
the RTT relations of Faddeev-Reshetikhin and Takhtajan \cite{FRT90}, one indeed starts with an $R$-matrix and
(if it satisfies the YBE) constructs a quasitriangular Hopf algebra with $R$ as its quasitriangular structure.

A special case of quasitriangular structure is one which satisfies $R_{21}=R^{-1}$. This is called a triangular
structure and the corresponding Hopf algebra is the one closest to our usual Lie algebraic notions. Any $R_F$ 
structure arising from a Hopf twist from the trivial quasitriangular structure $I\otimes I$ 
is automatically triangular, since $R_F=F_{21}(I\otimes I)F^{-1}_{12}$ implies that
\be
R_{F,21} R_{F,12}=F_{12}F^{-1}_{21}F_{21}F^{-1}_{12}=I\otimes I\;.
\ee
As shown by Drinfeld \cite{Drinfeld90}, the concept of a twist can be usefully 
extended to invertible elements of $\Hcal\otimes \Hcal$ which do \emph{not} satisfy the cocycle
conditions (\ref{cocyclecondition}). The resulting $R_F$-structure is then \emph{not} quasitriangular and 
the twisted coproduct $\Delta_F$ is \emph{not} associative. Since coassociativity is a defining property of 
Hopf algebras, the algebras one obtains in these way are in a more general class, called \emph{quasi-Hopf}
algebras. In a very similar way to quasitriangular Hopf algebras, which are non-cocommutative but the lack of cocommutativity
is controlled by the $R$-matrix (cf (\ref{quasitriangularR}), in quasi-Hopf algebras the lack of
coassociativity is controlled by an invertible element of $\Hcal\otimes \Hcal \otimes \Hcal$ 
called the \emph{coassociator}.

The relevance of quasi-Hopf algebras in our context is that it is likely that if, unlike \cite{Mansson:2008xv}, we
do not impose associativity on the twisted coproduct, the algebraic structure behind
the generic $(q,h)$ deformation will turn out to be a quasi-Hopf algebra. However, in this work we focus exclusively 
on special cases where the twist satisfies the cocycle condition and is thus a quasitriangular (and actually 
triangular, since we are twisting the trivial $R$-matrix $I\otimes I$) Hopf algebra.

\section{The hypergoniometric functions} \label{Hypergoniometric}

The expressions of the $w$-deformed $R$-matrix and twist are greatly simplified if one makes
use of the hypergoniometric sine and cosine functions $S(x)$ and $C(x)$. The former is 
defined through the integral \cite{Lundberg,LindkvistPeetre01}:
\be
x=\int^y \frac{\diff y}{(1-y^n)^{\frac mn}}
\ee
and inverting so that $y=S(x)$.  $S(x)$ is called the hypergoniometric sine (or \emph{sinualis}) of
order $m$ and degree $n$, sometimes denoted $S_\frac{m}{n}(x)$. (In this notation, the usual
trigonometric sine is $\sin(x)=S_\half(x)$, as can easily be checked). In our context, the relevant 
values are $m=1, n=3$, so in the text  $S(x)$ denotes $S_\frac{1}{3}(x)$. A description
of its properties can be found in \cite{Hypergoniometric}. 

The hypergoniometric cosine is defined through
\be
x=-\int_1^z\frac{\diff z}{(1-y^n)^{\frac mn}}
\ee
and inverting so that $z=C(x)$. We will again suppress the order and degree and 
 write $C(x)$ instead of the more accurate $C_{\frac{1}{3}}(x)$. 
 
The $S(x)$ and $C(x)$ functions satisfy relations that are completely analogous
to the usual trigonometric identities. In particular, for $m=1,n=3$ we have:
\be
S(x)^3+C(x)^3=1\;,\quad S(x)'=C(x)\quad\text{and}\quad C'(x)=-S(x)^2/C(x)\;.
\ee
For concreteness, let us record the first few terms in the series expansions of these functions (see also \cite{Hypergoniometric}):
\be
S(x) = x -2\frac{x^4}{4!} - 20\frac{x^7}{7!} - 3320\frac{x^{10}}{10!}+\cdots, \;\quad 
C(x) = 1 - 2\frac{x^3}{3!} - 20\frac{x^6}{6!} - 3320\frac{x^9}{9!}+\cdots
\ee
The reason that these functions make an appearance here is the nilpotency of our shift matrices $U$ and $V$. 
It can be straightforwardly checked that, for any $X$ such that $X^3=1$, and defining $b$ to 
be $b(a)=-\int^a\diff a S(a)/C(a)$,
we have
\be
e^{a X +b X^2}=C(a)+S(a)\cdot X\;.
\ee
That is, no terms proportional to $X^2$ appear in the series expansion. 
In section \ref{wDeformation}, this property was used to exponentialise the twist $F_w$, which 
greatly simplifies some manipulations, such as the action of the coproduct.

\section{Higher powers of the star product} \label{HigherStarProducts}

Let us briefly discuss how the star product defined through the twisted product (\ref{starproduct})
extends to higher powers. First, consider the cubic product:
\be
\begin{split}
(x\star y) \star z&=m_F(m_F(x\otimes y)\otimes z)=m_F\left((F^{-1}_{(1')} \trr x)(F^{-1}_{(2')}y) \otimes z\right)\\
&=m\left(F^{-1}_{(1)}\trr m(F^{-1}_{(1')}\trr x \otimes F^{-1}_{(2')}\trr y) \otimes F^{-1}_{(2)}\trr z\right)\\
&=m\left(\Delta(F^{-1}_{(1)})\trr [F^{-1}_{(1')}\trr x \otimes F^{-1}_{(2')}\trr y] \otimes F^{-1}_{(2)}\trr z\right)\\
&=m\left([F^{-1}_{(1)(1)}F^{-1}_{(1')}\otimes F^{-1}_{(1)(2)}F^{-1}_{(2')}\otimes F^{-1}_{(2)}]\trr [x\otimes y\otimes z]\right)\\
&=m\left((\Delta \otimes \id)(F^{-1})(F^{-1}\otimes 1)\trr [x\otimes y\otimes z]\right)\;.
\end{split}
\ee
Similarly, we compute
\be
\begin{split}
x\star (y\star z)&=m_F(x\otimes m_F(y\otimes z))=m_F(x\otimes (F^{-1}_{(1)}\trr y)(F^{-1}_{(2)}z))\\
&=m\left(F^{-1}_{(1)}\trr x \otimes F^{-1}_{(2)}\trr m(F^{-1}_{(1')}\trr y\otimes F^{-1}_{(2')}\trr z)\right)\\
&=m\left(F^{-1}_{(1)}\trr x \otimes \Delta(F^{-1}_{(2)})\trr [F^{-1}_{(1')}\trr y\otimes F^{-1}_{(2')}\trr z]\right)\\
&=m\left(F^{-1}_{(1)}\trr x \otimes F^{-1}_{(2)(1)}F^{-1}_{(1')}\trr y\otimes F^{-1}_{(2)(2)}F^{-1}_{(2')}\trr z\right)\\
&=m\left((\id\otimes \Delta) F^{-1}(1\otimes F^{-1})\trr [x\otimes y \otimes z]\right)\;.
\end{split}
\ee
(We have used the associativity of the untwisted module product to extend it as 
$m: C\otimes C\otimes C\ra C$, meaning just $m(X\otimes Y\otimes Z)= m(X\otimes m(Y\otimes Z))$)

We see that the cocycle condition guarantees the associativity of the twisted product. In our specific
case we can go further, since the twist satisfies the YBE. 
In that case we can write 
\be \label{tripletwists}
\begin{split}
(x\star y)\star z=m\left((F^{-1}_{23} F^{-1}_{13} F^{-1}_{12})\trr [x\otimes y \otimes z]\right)\\
x\star (y\star z)=m\left((F^{-1}_{12} F^{-1}_{13} F^{-1}_{23})\trr [x\otimes y \otimes z]\right)
\end{split}
\ee
which makes the relation between YBE and associativity even more direct. We emphasise, however,
that the weaker cocycle condition is sufficient for associativity (this has also been stressed in \cite{Watts:2000mq}). 
We conclude that we only need consider one of the above orderings in defining the star product. We pick:
\be
x\star y\star z=m\left((F^{-1}_{12} F^{-1}_{13} F^{-1}_{23})\trr [x\otimes y \otimes z]\right)
\ee
For many applications it is useful to also express this relation in index form, acting on the coordinates $z^i$. 
It is straightforward to check that it is equal to
\be
z^i\star z^j\star z^k=(F^{-1})^{ji}_{\;j'i'} (F^{-1})^{k l'}_{\;k'l} (F^{-1})^{k'j'}_{\;nm} z^l z^m z^n
\ee
where we note that an inversion in indices is required. 

Acting on four coordinates we obtain:
\be
((z^1\star z^2)\star z^3)\star z^4=m(F^{-1}_{34}F^{-1}_{24}F^{-1}_{14} F^{-1}_{23}F^{-1}_{13}F^{-1}_{12}\triangleright z^1\otimes z^2\otimes z^3\otimes z^4)\;,
\ee
\be
(z^1\star z^2)\star (z^3\star z^4)=m(F^{-1}_{23}F^{-1}_{13}F^{-1}_{24} F^{-1}_{14}F^{-1}_{12}F^{-1}_{34}\triangleright z^1\otimes z^2\otimes z^3\otimes z^4)
\ee
and so on for the various different placements of parentheses. Here we used the special 
properties of our twist to write the
final answer. We can again confirm that different placements of parentheses are equal, by manipulating
the twist matrices using the YBE. 

\section{The $w$-deformed generalised complex structures} \label{GCS}

In this appendix we record the generalised complex structures of the $w$-deformed background.
The generalised complex structure corresponding to the $\Phi_-$ pure spinor can be written as
\be
\Jcal_-{}^M_{\;N}=\twobytwo{J^{(ul)}_-}{J^{(ur)}_-}{0}{-J^{(ul)}_-}
\ee
where
\be
J^{(ul)}_-=\begin{pmatrix}-i&0&0&0&0&0\cr 0&i&0&0&0&0\cr 0&0&-i&0&0&0\cr 0&0&0&i&0&0
 \cr 0&0&0&0&-i&0\cr 0&0&0&0&0&i\cr \end{pmatrix}
\ee
and
\scriptsize
\be
J^{(ur)}_-=\begin{pmatrix}0&0&w\,\left({  z_3}^2-{  z_1}\,{  z_2}\right)&0&w\,
 \left({  z_1}\,{  z_3}-{  z_2}^2\right)&0\cr 0&0&0&w\,\left(
 {  \zb_3}^2-{  \zb_1}\,{  \zb_2}\right)&0&w\,\left({  \zb_1}\,
 {  \zb_3}-{  \zb_2}^2\right)\cr -w\,\left({  z_3}^2-{  z_1}\,
 {  z_2}\right)&0&0&0&-w\,\left({  z_2}\,{  z_3}-{  z_1}^2
 \right)&0\cr 0&-w\,\left({  \zb_3}^2-{  \zb_1}\,{  \zb_2}\right)&0&0
 &0&-w\,\left({  \zb_2}\,{  \zb_3}-{  \zb_1}^2\right)\cr -w\,\left(
 {  z_1}\,{  z_3}-{  z_2}^2\right)&0&w\,\left({  z_2}\,
 {  z_3}-{  z_1}^2\right)&0&0&0\cr 0&-w\,\left({  \zb_1}\,
 {  \zb_3}-{  \zb_2}^2\right)&0&w\,\left({  \zb_2}\,{  \zb_3}-
 {  \zb_1}^2\right)&0&0\cr \end{pmatrix}\;.
\ee
\normalsize
The generalised complex structure corresponding to $\Phi_+$ can be written as
\be
\Jcal_+{}^M_{\;N}=\twobytwo{J^{(ul)}_+}{J^{(ur)}_+}{(J^{(ur)}_+)^T}{-(J^{(ul)}_+)^T}
\ee
with
\scriptsize
\be
J^{(ul)}_+=\left(\begin{array}{cccccc}
{{w\,\left({  z_3}\,{  \zb_3}-{  z_2}\,{  \zb_2}
 \right)}\over{2}}&0&-{{w\,\left({  z_2}\,{  \zb_3}-{  \zb_1}\,
 {  z_3}\right)}\over{2}}&{{w\,\left({  z_3}^2-{  z_1}\,
 {  z_2}\right)}\over{2}}&{{w\,\left({  \zb_2}\,{  z_3}-{  \zb_1}
 \,{  z_2}\right)}\over{2}}&{{w\,\left({  z_1}\,{  z_3}-
 {  z_2}^2\right)}\over{2}}\cr 0&{{w\,\left({  z_3}\,{  \zb_3}-
 {  z_2}\,{  \zb_2}\right)}\over{2}}&{{w\,\left({  \zb_3}^2-
 {  \zb_1}\,{  \zb_2}\right)}\over{2}}&{{w\,\left({  z_1}\,
 {  \zb_3}-{  \zb_2}\,{  z_3}\right)}\over{2}}&{{w\,\left({  \zb_1}
 \,{  \zb_3}-{  \zb_2}^2\right)}\over{2}}&{{w\,\left({  z_2}\,
 {  \zb_3}-{  z_1}\,{  \zb_2}\right)}\over{2}}\cr {{w\,\left(
 {  z_1}\,{  \zb_3}-{  \zb_2}\,{  z_3}\right)}\over{2}}&-{{w\,
 \left({  z_3}^2-{  z_1}\,{  z_2}\right)}\over{2}}&-{{w\,\left(
 {  z_3}\,{  \zb_3}-{  z_1}\,{  \zb_1}\right)}\over{2}}&0&-{{w\,
 \left({  \zb_1}\,{  z_3}-{  z_1}\,{  \zb_2}\right)}\over{2}}&-{{
 w\,\left({  z_2}\,{  z_3}-{  z_1}^2\right)}\over{2}}\cr -{{w\,
 \left({  \zb_3}^2-{  \zb_1}\,{  \zb_2}\right)}\over{2}}&-{{w\,\left(
 {  z_2}\,{  \zb_3}-{  \zb_1}\,{  z_3}\right)}\over{2}}&0&-{{w\,
 \left({  z_3}\,{  \zb_3}-{  z_1}\,{  \zb_1}\right)}\over{2}}&-{{
 w\,\left({  \zb_2}\,{  \zb_3}-{  \zb_1}^2\right)}\over{2}}&-{{w\,
 \left({  z_1}\,{  \zb_3}-{  \zb_1}\,{  z_2}\right)}\over{2}}\cr 
 {{w\,\left({  z_2}\,{  \zb_3}-{  z_1}\,{  \zb_2}\right)}\over{2
 }}&-{{w\,\left({  z_1}\,{  z_3}-{  z_2}^2\right)}\over{2}}&-{{
 w\,\left({  z_1}\,{  \zb_3}-{  \zb_1}\,{  z_2}\right)}\over{2}}&
 {{w\,\left({  z_2}\,{  z_3}-{  z_1}^2\right)}\over{2}}&{{w\,
 \left({  z_2}\,{  \zb_2}-{  z_1}\,{  \zb_1}\right)}\over{2}}&0
 \cr -{{w\,\left({  \zb_1}\,{  \zb_3}-{  \zb_2}^2\right)}\over{2}}&{{
 w\,\left({  \zb_2}\,{  z_3}-{  \zb_1}\,{  z_2}\right)}\over{2}}&
 {{w\,\left({  \zb_2}\,{  \zb_3}-{  \zb_1}^2\right)}\over{2}}&-{{w\,
 \left({  \zb_1}\,{  z_3}-{  z_1}\,{  \zb_2}\right)}\over{2}}&0&
 {{w\,\left({  z_2}\,{  \zb_2}-{  z_1}\,{  \zb_1}\right)}\over{2
 }}\cr \end{array}\right)
\ee
\normalsize
and
\be
J^{(ur)}_+=\begin{pmatrix}0&-i&0&0&0&0\cr i&0&0&0&0&0\cr 0&0&0&-i&0&0\cr 0&0&i&0&0&0
 \cr 0&0&0&0&0&-i\cr 0&0&0&0&i&0\cr \end{pmatrix}
\ee
These generalised complex structures both square to $-1$ and commute. They thus define 
a generalised complex metric via (\ref{generalisedmetricdef}). 

\section{The IIB supergravity equations} \label{IIBEOM}

For completeness, in this appendix we write down the field equations of IIB supergravity that our solutions
satisfy. Restricting at first to the NS-NS sector, the relevant equations are as follows \cite{Schwarz:1983qr,Gauntlett:2005ww}:\footnote{In this appendix $M,N,\ldots$ denote the coordinates of ten-dimensional spacetime.}
\be
R_{MN}=\half\p_M\Phi\p_N\Phi+\frac14 e^{-\Phi}H_{MRS}H_{N}^{\;\;RS}-\frac{1}{48}e^{-\Phi} g_{MN}H_{RST}H^{RST}
\ee
where $H=dB$\;, 
\be
\half\nabla^M\p_M\Phi=-\frac1{24}\frac{1}{\sqrt{G}}H_{MNR}H^{MNR}\;.
\ee
and
\be
D^P\left(e^{-\Phi/2} H_{MNP}\right)=\half (D^P \Phi) e^{-\Phi/2} H_{MNP}\;.
\ee
The solution exhibited in section \ref{Precursor} can be straightforwardly shown to satisfy these equations.

For the RR solution, we prefer to write the IIB equations in string frame ($g^{(s)}_{MN}=e^{\Phi/2}g^{(E)}_{MN}$), where
the 10-d metric factorises into an AdS$_5$ and $S_w^5$ part. We also take the following ansatz for the five-form field
strength $F=\diff C_4$:
\be
F_{(5)}=\omega_{AdS_5}+\omega_{S^5_w}
\ee
with 
\be
\omega_{S^5_w}=R^5\sqrt{g}\diff\alpha\diff\theta\diff\phi_1\diff\phi_2\diff\phi_3=
G  \sin^3\alpha\cos\alpha\sin\theta\cos\theta
\ee
ensuring self-duality. We also set the RR scalar $C_{0}$ to zero. 

There are two three-form equations of motion. The first one is
\be
F_{MNP}=-\frac{R}{24} D_M \sqrt{g}e^{-2\Phi}\epsilon_{NPQRS}H^{QRS}
\ee
which (knowing the B-field) we use to obtain the RR three-form field strength. This of course ensures that it is satisfied. 
It is also compatible with the other three-form equation of motion
\be
H_{MNP}=\frac{R}{24} D_M \sqrt{g}\epsilon_{NPQRS}H^{QRS}
\ee
Thus the two equations to be checked are the Einstein equations:
\be
\begin{split}
R_{MN}&=-2D_M\p_N\Phi-\frac14g_{MN}D^P\p_P\Phi+\half g_{MN}\p_P\Phi\p^P\Phi+\frac{1}{96}e^{2\Phi}F_{MPQRS}F_{N}^{\;\;PQRS}\\
&\quad+\frac14\left(H_{MPQ}H_N^{\;\;PQ}+e^{2\Phi}F_{MPQ}F_{N}^{\;\;PQ}\right)-\frac{1}{48}g_{MN}\left(H_{MNP}H^{MNP}+e^{2\Phi} F_{MNP}F^{MNP}\right)
\end{split}
\ee
and the dilaton equation:
\be
D^M\p_M e^{-2\Phi}=-\frac16\left(F_{MNP}F^{MNP}-e^{-2\Phi}H_{MNP}H^{MNP}\right)
\ee
The task of checking these equations for the solution in section \ref{RRsolution} is formidable,
given the complexity of the metric in the coordinate system used. As discussed in the text, the
solution has been verified as an expansion in $w$ (up to and including $O(w^6)$) as well
as exactly in $w$ but for specific choices of angles. Of course, a coordinate redefinition to the
standard LM coordinates brings the $w$-deformed metric into a far more manageable form.

\bibliography{HopfGG}
\bibliographystyle{utphys}

\end{document}